\documentclass[sn-aps]{sn-jnl}

\jyear{2021}%

\theoremstyle{thmstyleone}%
\theoremstyle{thmstyletwo}%

\theoremstyle{thmstylethree}%

\begin{document}

\title[A Generalized Feynman-Kac Approach to Subdiffusion with Reactions]{Subdiffusion in the Presence of Reactive Boundaries: A Generalized Feynman-Kac Approach}


\author*[1]{\fnm{Toby} \sur{Kay}}\email{toby.kay@bristol.ac.uk}

\author[1,2]{\fnm{Luca} \sur{Giuggioli}}

\affil*[1]{\orgdiv{Department of Engineering Mathematics}, \orgname{University of Bristol}, \orgaddress{\city{Bristol}, \postcode{BS8 1UB}, \country{United Kingdom}}}

\affil[2]{\orgdiv{Bristol Centre for Complexity Sciences}, \orgname{University of Bristol}, \orgaddress{\city{Bristol}, \postcode{BS8 1UB}, \country{United Kingdom}}}

\abstract{We derive, through subordination techniques, a generalized Feynman-Kac equation in the form of a time fractional Schr\"{o}dinger equation. We relate such equation to a functional which we name the subordinated local time. We demonstrate through a stochastic treatment how this generalized Feynman-Kac equation describes subdiffusive processes with reactions. In this interpretation, the subordinated local time represents the number of times a specific spatial point is reached, with the amount of time spent there being immaterial. This distinction provides a practical advance due to the potential long waiting time nature of subdiffusive processes. The subordinated local time is used to formulate a probabilistic understanding of subdiffusion with reactions, leading to the well known radiation boundary condition. We demonstrate the equivalence between the generalized Feynman-Kac equation with a reflecting boundary and the fractional diffusion equation with a radiation boundary. We solve the former and find the first-reaction probability density in analytic form in the time domain, in terms of the Wright function. We are also able to find the survival probability and subordinated local time density analytically. These results are validated by stochastic simulations that use the subordinated local time description of subdiffusion in the presence of reactions.}


\maketitle

\section{Introduction}\label{sec:introduction}

In recent years anomalous diffusion has been found to be a prevalent transport mechanism across many systems. Specifically, subdiffusion is of key interest due to its defining sub-linear mean-square displacement in time, i.e. 
\begin{equation*}
    \langle Y^2(t) \rangle \sim t^\alpha, 
\end{equation*}
where $Y(t)$ is a time dependent random variable with $\alpha \in (0,1)$. Due to this sub-linear form, subdiffusive motion has been observed in a variety of physical and biological processes (see \cite{metzler2000random,metzler2004restaurant} and references therein). Over the last two decades or so, much work has been endeavored to create a unified framework to describe subdiffusive motion. One of the most utilised approach is a fractional diffusion or Fokker-Planck equation, derived from a generalized master equation (GME) or continuous-time random walk (CTRW) approach \cite{metzler1999deriving,barkai2000continuous,giuggioli2009generalized}. It is also possible to obtain this fractional diffusion equation through a subordinated Langevin approach \cite{fogedby1994langevin}.

Within this unified framework it is natural to consider another fundamental equation in the study of stochastic processes, the Feynman-Kac equation (FKE), and its extension to the subdiffusive case. The classical FKE is a well known tool to study functionals of Brownian motion with numerous applications across physics and other areas of science \cite{majumdar2007brownian}. Thus there was a clear need for the extension to when the underlying stochastic process is subdiffusive. This need has been met in recent years with a fractional FKE having been found to study functionals of subdiffusion \cite{turgeman2009fractional,carmi2010distributions,carmi2011fractional,cairoli2015anomalous} with further generalizations to space and time dependent forces \cite{zhang2013fractional,cairoli2017feynman}, tempered subdiffusion \cite{wu2016tempered}, aging subdiffusion \cite{wang2017aging}, multiplicative noise \cite{wang2018feynman} and reaction-subdiffusion processes \cite{hou2018feynman}. 

One of the most utilized functionals is the so-called local time functional \cite{levy1940certains}, which finds applications in various areas \cite{majumdar2002local}, and has recently been used to build a probabilistic description of diffusion with surface reactions \cite{grebenkov2019probability,grebenkov2020imperfect,grebenkov2020paradigm}. In this approach surface reactions are described via stopping conditions based on the local time of the Brownian particle at the boundary. The Brownian particle undergoes normal diffusion being reflected each time it reaches the boundary until the local time at the boundary exceeds a random variable drawn from an exponential distribution with the inverse scale being the reactivity parameter. The time at which this occurs is then the reaction time, where the particle has then reacted, absorbed, changed species etc. This approach presents a formal and practical advance compared to classical methods such as using radiation boundary conditions \cite{redner2001guide} or placing partial traps or defects in the domain \cite{szabo1984localized,kay2022defect}.

Due to the ubiquity of subdiffusion in complex systems, these systems are often bounded by reactive boundaries \cite{grebenkov2010subdiffusion}. Thus there is a clear need to generalize this description for when the motion is subdiffusive. The main purpose of this paper is to provide such a generalization. We do this by considering an alternative generalized FKE \cite{magdziarz2015asymptotic,bender2022subordination} rather than the previously mentioned fractional FKE. The generalized FKE we use is in the form of an (imaginary time) time fractional Schr\"{o}dinger equation \cite{naber2004time,iomin2009fractional,achar2013time,bayin2013time} and governs subordinated forms of the functionals. This proves to be a useful recipe in the case of the local time functional for providing such a generalized description of subdiffusion in the presence of reactive boundaries.

The paper is structured as follows. In Sec. \ref{sec:frac_fk} we recall the classical FKE to which we derive a generalized form through subordination techniques and introduce the subordinated local time functional whose meaning is uncovered using a CTRW approach. In Sec. \ref{sec:reactive_boundaries} we present a probabilistic interpretation of this generalized FKE as subdiffusion with reactions using the subordinated local time and how this is connected to the radiation boundary condition (BC). In Sec. \ref{sec:first_reac_time} we present an application of these findings by analytically studying three important quantities associated with subdiffusion in the presence of a radiation boundary, namely the first-reaction time density, survival probability and subordinated local time density. We confirm these analytic results with stochastic simulations. Finally, we discuss and conclude our findings in Sec. \ref{sec:conclusion}.

\section{A Generalized Feynman-Kac Equation}\label{sec:frac_fk}

\subsection{The Classical Feynman-Kac Equation}

The celebrated FKE, derived in 1949 by Kac influenced by Feynman's path integral description of quantum mechanics, has become a fundamental tool in the theory of stochastic processes \cite{kac1949distributions,kac1951some}. The Feynman-Kac theory provides a rigorous connection between the paths, $X(t)$, of a Bronwnian motion process and the solution to the (imaginary time) Schr\"{o}dinger equation \cite{oksendal2003stochastic}. The main utility however is the connection to functionals of Brownian motion \cite{majumdar2007brownian},
\begin{equation}\label{eq:functional}
    \mathcal{A}(t)=\int_0^t U[X(t')]dt',
\end{equation}
where $U(x)$ is some arbitrary function. The FKE governs the (Laplace/Fourier transformed) joint probability density, $\rho(x,A,t\vert x_0)$, of $X(t)$ and $\mathcal{A}(t)$, given by \cite{oksendal2003stochastic,schuss2015brownian}
\begin{equation}\label{eq:normal_fk}
    \frac{\partial }{\partial t} P(x,p,t\vert x_0)=K \frac{\partial^2}{\partial x^2} P(x,p,t\vert x_0)-p U(x) P(x,p,t\vert x_0),
\end{equation}
where $K$ is the diffusion coefficient, i.e. it is the strength of the delta correlated noise of the Langevin equation associated with $X(t)$, while the Laplace variable $p$ is related to $A$ via \cite{majumdar2007brownian}
\begin{equation}\label{eq:normal_fk_sol_laplace}
    P(x,p,t\vert x_0)=\int_0^\infty e^{-p A} \rho(x,A,t\vert x_0) dA.
\end{equation}
It should be noted that if $\mathcal{A}(t)$ is not always positive, then the Laplace transform needs to be replaced by a Fourier transform, i.e. $p \to -i p$ and the lower integration bound changed to $-\infty$ \cite{carmi2010distributions}. Alternatively, Eq. (\ref{eq:normal_fk_sol_laplace}) can be represented via the expectation,
\begin{equation}\label{eq:normal_fk_sol}
    P(x,p,t\vert x_0)=\left\langle \delta(X(t)-x) e^{-p \int_0^t U[X(t')]dt'} \right\rangle_{x_0}.
\end{equation}
In Eq. (\ref{eq:normal_fk_sol}) the average is over all trajectory realizations of $X(t)$ that starts at $X(0)=x_0$, that is $P(x,p,0\vert x_0)=\delta(x-x_0)$.

\subsection{Time-Changed Process}

Let us now consider subdiffusion through a CTRW paradigm \cite{montroll1965random,weiss1994aspects}. A CTRW formalism is constructed by considering a random walker which waits at each step, $i$, for a time $\eta_i$ and then proceeds to jump a distance $\xi_i$. The random variables $\xi_i$ and $\eta_i$, are independent and identically distributed. Thus, after $n$ steps, the position of the random walker, $Y_n$ and the total time elapsed, $T_n$, can be found by \cite{cairoli2017feynman},
\begin{equation}\label{eq:discrete_ctrw}
    Y_n=Y_0+\sum_{i=1}^n \xi_i \quad \mbox{and} \quad T_n=\sum_{i=1}^n\eta_i,
\end{equation}
where $Y_0$ is the initial position. Through a parameterization of the CTRW via the continuous time variable, $t$, instead of the number of steps, $n$, Eq. (\ref{eq:discrete_ctrw}) can be written compactly as,
\begin{equation}\label{eq:ctrw_param}
    Y(t)=Y_0+\sum_{i=1}^{N(t)}\xi_i,
\end{equation}
where $N(t)=\max\left\{n\geq0:T_n\leq t\right\}$, such that $N(t)$ is a random variable itself, as a consequence of containing the statistics of the random waiting times. If we now take the continuum limit of Eq. (\ref{eq:discrete_ctrw}), we obtain \cite{cairoli2017feynman}
\begin{equation}\label{eq:langevin}
    X(\tau)=X_0+\int_0^\tau \xi(s)ds \quad \mbox{and} \quad T(\tau)=\int_0^\tau \eta(s)ds.
\end{equation}
Here, $\tau$ is not the real physical time, but is instead an operational time. 

In the same sense, we take the continuum limit of Eq. (\ref{eq:ctrw_param}), with $N(t)\to S(t)$, to obtain
\begin{equation}
    Y(t)=Y_0+\int_0^{S(t)} \xi(s)ds.
\end{equation}
Since $S(0)=0$, we have $Y_0=X_0=x_0$, therefore $Y(t)=X(S(t))$, with $S(t)=\inf\{\tau>0:T(\tau)>t\}$. In other words $Y(t)$ has undergone a time change and is a subordinated process, such that $S(t)$ can be interpreted as a stochastic clock \cite{fogedby1994langevin}. 

Specifically, subdiffusion is generated in the macroscopic limit of a CTRW with a distribution of waiting times that is heavy-tailed, such that the mean waiting time is infinite. If we indicate with $T_\alpha(\tau)$ a waiting time distribution which follows an $\alpha$-stble L\'{e}vy distribution, the Laplace transform of $T_\alpha(\tau)$ is $\langle \exp\{-\kappa T_\alpha(\tau) \} \rangle= \exp\{- \tau \kappa^\alpha \}$, with $\alpha\in(0,1)$ \cite{janicki2021simulation}. The stochastic clock is then defined as, 
\begin{equation}\label{eq:subordinator_def}
    S_\alpha(t)=\inf\{\tau>0:T_\alpha(\tau)>t\}, 
\end{equation}
and will be termed the inverse $\alpha$-stable subordinator \cite{meerschaert2002stochastic,piryatinska2005models,magdziarz2008equivalence}. The Laplace transform of the probability density of $S_\alpha(t)$, $g(\tau,t)$, is given by \cite{baule2005joint}
\begin{equation}\label{eq:inverse_sub_density}
    \widetilde{g}(\tau,\epsilon)= \int_0^\infty e^{-\epsilon t} g(\tau,t) dt =\epsilon^{\alpha-1}e^{-\tau \epsilon^\alpha},
\end{equation}
alternatively by taking the derivative of both sides of Eq. (\ref{eq:inverse_sub_density}) and performing the inverse Laplace transform, one can show $g(\tau,t)$ satisfies the following fractional differential equation \cite{baule2005joint},
\begin{equation}\label{eq:frac_diff_eq}
    \frac{\partial g(\tau,t)}{\partial t}=-{}_0D_t^{1-\alpha}\frac{\partial g(\tau,t)}{\partial \tau}.
\end{equation}
Here ${}_0D_t^{1-\alpha}$ is a fractional derivative of Riemann-Liouville type \cite{podlubny1999fractional}, i.e. for a generic function $f(t)$,
\begin{equation}
    {}_0D_t^{1-\alpha} f(t)=\frac{1}{\Gamma(\alpha)}\frac{\partial}{\partial t} \int_{0}^{t} \frac{f(t')}{(t-t')^{1-\alpha}}dt'.
\end{equation}

\begin{figure}[h!]
    \centering
    \includegraphics[width=1 \textwidth]{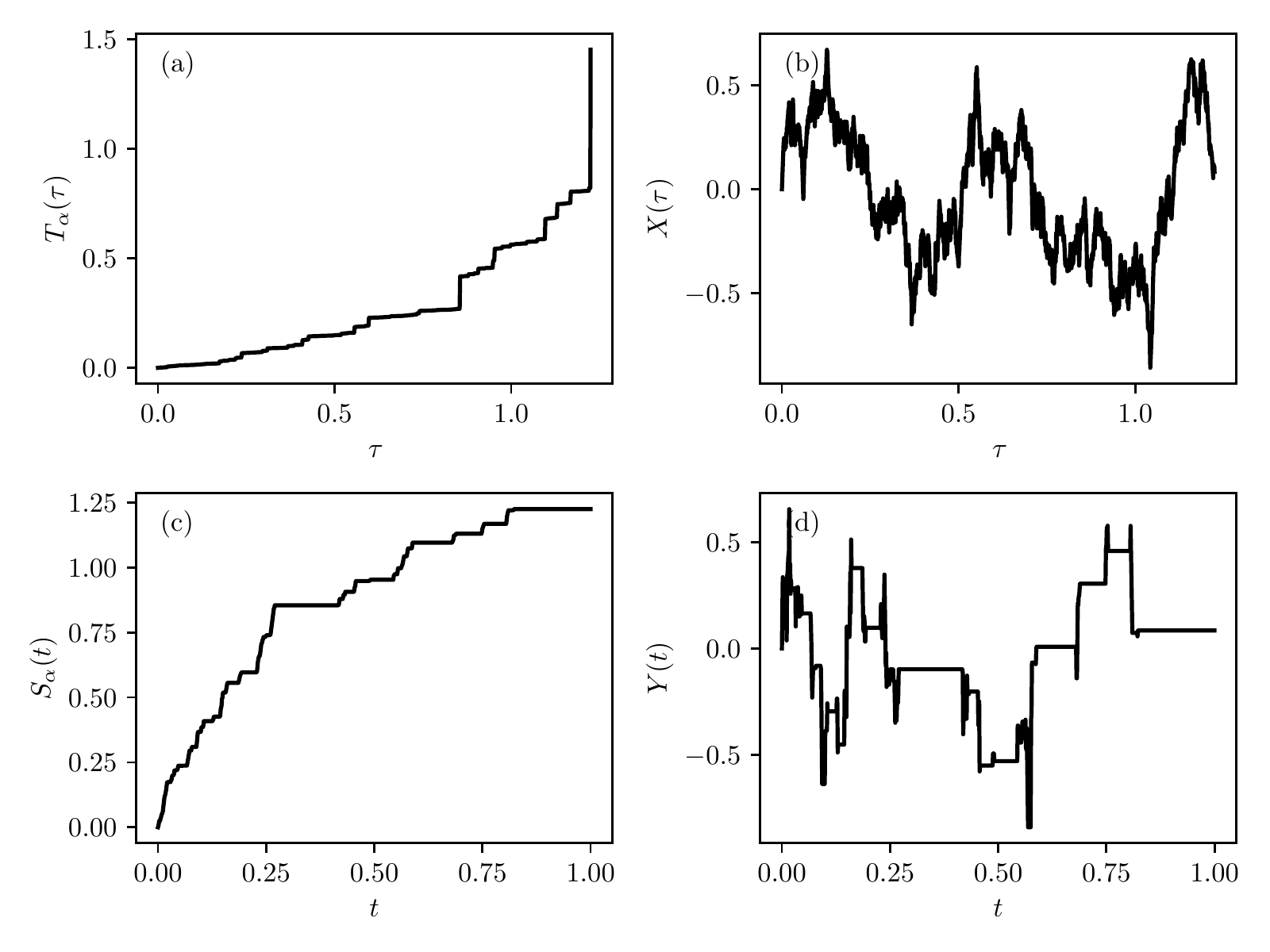}
    \caption{A set of trajectories for the four process that leads to subdiffusion, (a) $\alpha$-stable L\'{e}vy motion, (b) Brownian motion, (c) inverse $\alpha$-stable subordinator, (d) subdiffusion. (a) and (b) are both in terms of the operational time $\tau$ and (c) and (d) are in terms of the real physical time $t$. These plots were generated using the algorithm in \cite{magdziarz2007fractional}, where $T_\alpha(\tau)$ is simulated to which $S_\alpha(t)$ is found using Eq. (\ref{eq:subordinator_def}). Then $Y(t)=X(S_\alpha(t))$ can be approximated by interpolating $X(\tau)$ for the values of $S_\alpha(t)$. One can see that the delays in (c) correspond to periods of waiting in (d).}
    \label{fig:subdiffusion_subordinate_sim}
\end{figure}

When $\xi(\tau)$ is the standard Langevin force (i.e. Gaussian white noise), $Y(t)=X(S_\alpha(t))$ is a subdiffusion process, with $X(\tau)$ being standard Brownian motion with probability density, $W(x,\tau \vert x_0)=P(x,0,\tau\vert x_0)$. In Fig. (\ref{fig:subdiffusion_subordinate_sim}) we show a simple realization of $T_\alpha(\tau)$ and its corresponding $S_\alpha(t)$, and how the resulting $X(\tau)$ trajectory is modified to a $Y(t)$ trajectory. 

The probability density of $Y(t)$ written as $W_\alpha(x,t\vert x_0)$ can be given in terms of $W(x,\tau\vert x_0)$ and $g(\tau,t)$ as follows \cite{sokolov2000levy}, 
\begin{equation}\label{eq:total_prob}
    W_\alpha(x,t\vert x_0)=\int_0^\infty W(x,\tau\vert x_0) g(\tau,t)d\tau,
\end{equation}
due to the independence of $X(\tau)$ and $S_\alpha(t)$. By taking the time derivative of both sides of Eq. (\ref{eq:total_prob}), whilst using Eq. (\ref{eq:frac_diff_eq}) and integrating by parts we find, $\partial_t W_\alpha(x,t\vert x_0)=-{}_0D_t^{1-\alpha}\Big\{[g(\tau,t)W(x,\tau\vert x_0)]_0^\infty- \int_{0}^{\infty} g(\tau,t) \partial_\tau W(x,\tau\vert x_0)d\tau \Big\}$. From Eq. (\ref{eq:inverse_sub_density}) it is clear that $g(\infty,t)=0$ and ${}_0D_t^{1-\alpha} g(0,t)=\delta(t)$, thus for $t>0$, and using the normal diffusion equation, we recover the fractional diffusion equation (FDE) \cite{metzler2000random,barkai2001fractional,metzler2004restaurant},
\begin{equation}\label{eq:fractional_diffusion}
    \frac{\partial}{\partial t}W_\alpha(x,t\vert x_0) = K_\alpha \, {}_0D_t^{1-\alpha}\frac{\partial^2}{\partial x^2}W_\alpha(x,t\vert x_0),
\end{equation}
with $K_\alpha$ being the generalized diffusion coefficient which has dimensions $[\text{length}]^2/[\text{time}]^\alpha$.

Now let us consider not only subordinating the Brownian motion but also the functional of Brownian motion, i.e. 
\begin{equation}\label{eq:sub_functional}
    \mathcal{A}(S_\alpha(t))=\int_{0}^{S_\alpha(t)}U[X(t')]dt'.
\end{equation}
The joint density, $P_\alpha(x,p,t\vert x_0)$, where
\begin{equation}\label{eq:total_prob_fk}
    P_\alpha(x,p,t\vert x_0)=\int_0^\infty g(\tau,t) P(x,p,\tau\vert x_0) d\tau,
\end{equation}
will then be governed by a generalized FKE. Using the same arguments as above we have \cite{magdziarz2015asymptotic},
\begin{equation}\label{eq:fractional_fk}
    \frac{\partial}{\partial t}P_\alpha(x,p,t\vert x_0)={}_0D_t^{1-\alpha}\left[K_\alpha \frac{\partial^2}{\partial x^2} -p U(x)  \right]P_\alpha(x,p,t\vert x_0).
\end{equation}
It is then simple to find $P_\alpha(x,p,t\vert x_0)$ via Eq. (\ref{eq:total_prob_fk}), since
\begin{equation}
    P_\alpha(x,p,t\vert x_0)=\int_0^\infty \langle \delta(\tau-S_\alpha(t)) \rangle P(x,p,\tau\vert x_0) d\tau.
\end{equation}
Using Eq. (\ref{eq:normal_fk_sol}) and the properties of independence, we have
\begin{eqnarray}\label{eq:fractional_fk_sol}
     P_\alpha(x,p,t\vert x_0)=\left\langle \delta(X(S_\alpha(t))-x) e^{-p \int_0^{S_\alpha(t)} U[X(t')]dt'} \right\rangle_{x_0}\\ \nonumber
     =\int_0^\infty e^{-p A} \rho_\alpha(x,A,t\vert x_0) dA,
\end{eqnarray}
where $\rho_\alpha(x,A,t\vert x_0)$ is the joint density of $X(S_\alpha(t))$ and $\mathcal{A}(S_\alpha(t))$. 

We point out the difference of Eq. (\ref{eq:fractional_fk}) compared to the fractional FKE, as mentioned in Sec. \ref{sec:introduction}, which is given by \cite{turgeman2009fractional,carmi2010distributions},
\begin{equation}\label{eq:substantial_FK}
    \frac{\partial }{\partial t} P_\alpha(x,p,t\vert x_0)= K_\alpha \frac{\partial^2}{\partial x^2} \mathcal{D}_t^{1-\alpha} P_\alpha(x,p,t\vert x_0) -p U(x) P_\alpha(x,p,t\vert x_0)
\end{equation}
where $\mathcal{D}_t^{1-\alpha}$ is the so-called fractional substantial derivative \cite{friedrich2006anomalous},
\begin{equation}
    \mathcal{D}_t^{1-\alpha}P_\alpha(x,p,t\vert x_0) = \frac{1}{\Gamma(\alpha)}\left[\frac{\partial}{\partial t} + p U(x)\right] \int_{0}^{t} \frac{e^{-(t-t')pU(x)}}{(t-t')^{1-\alpha}} P_\alpha(x,p,t'\vert x_0)dt'.
\end{equation}
Thus the corresponding density, $\rho_\alpha(x,A,t\vert x_0)$, is then the joint density of $Y(t)$ and $\mathcal{A}_\alpha(t)=\int_{0}^{t} U[Y(t')]dt'$ (compare to Eq. (\ref{eq:sub_functional})). Thus the functional is of the subdiffusive motion and is not a subordinated quantity, illustrating that Eq. (\ref{eq:substantial_FK}) is the natural generalization of the FKE for studying functionals of subdiffusion.

\subsection{Subordinated Local Time}\label{sec:sub_local_time}

The local time of a stochastic process, originally introduced by L\'{e}vy \cite{levy1940certains}, is an important quantity that characterises the fraction of time the process spends at a certain point, $x_b$. We will label the local time, $\ell(t)$, which is defined as the functional $\mathcal{A}(t)$ with $U(x)=\delta(x-x_b)$, therefore
\begin{equation}\label{eq:local_time}
    \ell(t)=\int_0^t \delta(X(t')-x_b)dt'.
\end{equation}
Without loss of generality, moving forward, we take this point to be at the origin, $x_b=0$. 

Now, let us consider the subordinated local time, $\ell(S_\alpha(t))$, we are able to evince the meaning of this quantity, as follows. Returning to the CTRW paradigm (for the subdiffusive case), where the random walker is described by Eq. (\ref{eq:ctrw_param}) we introduce the quantity $\mathcal{N}(t)$, which is the number of times the walker visits a region around the origin, $\partial \Omega$, of width $\varepsilon$ ($\varepsilon$ gives the scale of the jump length i.e. $\langle \xi_i^2 \rangle=\varepsilon^2$), up to time $t$ \cite{it1965diffusion}:
\begin{equation}\label{eq:discrete_visits}
    \mathcal{N}(t)=\sum_{i=0}^{N(t)}\mathbb{I}_{\partial \Omega}(Y_i).
\end{equation}
In Eq. (\ref{eq:discrete_visits}) $Y_i$ is given by Eq. (\ref{eq:discrete_ctrw}) and $\mathbb{I}_{\partial \Omega}(Y_i)$ is the indicator function, i.e $\mathbb{I}_{\partial \Omega}(Y_i)=1$ if $Y_i\in \partial \Omega$ and $0$ otherwise. Now we take a continuum limit such that $Y_n \to X(\tau)$ and $N(t) \to S_\alpha(t)$ and introduce a scaling of $K_\alpha/\varepsilon^2$ to make $\mathcal{N}(t)$ dimensionless, 
\begin{equation}\label{eq:visits_lim}
    \mathcal{N}(t)=\frac{K_\alpha}{\varepsilon^2} \int_0^{S_\alpha(t)} \mathbb{I}_{\partial \Omega}(X(s))ds.
\end{equation}
Let us then take the diffusive limit, which entails letting $\varepsilon$ vanish, resulting in the indicator function, $\mathbb{I}_{\partial \Omega}(X(s))$, becoming the Dirac-$\delta$ function, then we have
\begin{equation}\label{eq:visits_macro}
    \lim_{\varepsilon \to 0} \varepsilon \mathcal{N}(t)=K_\alpha \ell(S_\alpha(t)).
\end{equation}
The limit in Eq. (\ref{eq:visits_macro}) exists due to the recurrent nature of Brownian motion in one dimension such that the number of visits, $\mathcal{N}(t)$, diverges in the diffusive limit and can be understood as the continuous analog of the (scaled) number of times a subdiffusive particle visits a certain point. Note this limit will not hold in higher dimensions, so one would need to consider a bounded domain where $\partial \Omega$ becomes a thin layer at a reflecting boundary \cite{grebenkov2019probability,grebenkov2020paradigm}. It is well known that the local time for a normal diffusive particle is the continuous limit of the number of times the particle visits a certain point \cite{it1965diffusion}. However, due to the long waiting times embedded in the subdiffusive dynamics a particle may spend anomalously long times at a certain point. Thus, the local time for a subdiffusive particle, $\ell_\alpha(t)=\int_0^t \delta(Y(t'))dt'$, does not correspond to the continuous limit of the number of times the particle visits a point. So, if one only cares about whether the particle has reached a certain point a number of times, and not how long it has spent there, e.g. if an event occurs when the particle reaches that point for every visit, then the quantity of interest is the subordinated local time, Eq. (\ref{eq:visits_macro}), i.e. $\ell(S_\alpha(t))=\int_0^{S_\alpha(t)} \delta(X(t'))dt'$. Note that $\ell(S_\alpha(t))$ has dimensions $[\text{time}]^\alpha/[\text{length}]$, so contrary to its name $\ell(S_\alpha(t))$ is not actually a time.

\section{Reactive Boundaries}\label{sec:reactive_boundaries}
\subsection{Generalized Feynman-Kac Equation and Reactions}\label{sec:frac_fk_react}
It is well known that the FKE (\ref{eq:normal_fk}) can be interpreted as diffusion with killing or reactions \cite{schuss2015brownian}, such that the diffusive particle is removed from the system i.e. by being absorbed. Let us say for simplicity that absorption occurs at the origin, thus we take $U(x)=\delta(x)$. We introduce the killing or first reaction time $\mathcal{T}$ and assume the reaction dynamics are subordinated and thus governed by the stochastic clock, $S_\alpha(t)$. Then the probability for a subdiffusive particle starting at $x_0$ to be killed in the time interval $[t,t+h)$ can be approximated by
\begin{eqnarray}\label{eq:approx_kill_prob}
    \mathbb{P}[\mathcal{T}\in[t,t+h)]_{x_0} \approx p \left\langle \delta(X(S_\alpha(t))) \right \rangle_{x_0}  (S_\alpha(t+h)-S_\alpha(t)).
\end{eqnarray}
Eq. (\ref{eq:approx_kill_prob}) can be understood as follows. Every time the particle reaches the origin there is a chance that a reaction occurs and the trajectory will be killed. As we are considering the reaction dynamics are occurring according to $S_\alpha(t)$, the physical time interval $[t,t+h)$ corresponds to the interval $[S_\alpha(t),S_\alpha(t+h))$ for the reaction dynamics. So for $h<<1$ one can approximate $\mathbb{P}[\mathcal{T}\in[t,t+h)]_{x_0}$ by multiplying the probability the particle is found at the origin at time $t$ having not reacted previously, $\left\langle \delta(X(S_\alpha(t))) \right \rangle_{x_0}$, by the interval over which the reactions occur, $(S_\alpha(t+h)-S_\alpha(t))$, and by the Laplace variable in Eq. (\ref{eq:fractional_fk_sol}), $p$, which here is considered as the reactivity. 

As each reaction event is taken to be independent of each other, the probability the particle has not reacted (survived) up to time t is given by,
\begin{eqnarray}\label{eq:survival_prod}
    \mathbb{P}[\mathcal{T}>t]_{x_0}    =\lim_{h \to 0} \prod_{i=1}^M \left[ 1-  p \left\langle \delta(X(S_\alpha(t_i))) \right \rangle_{x_0} (S_\alpha(t_i+h)-S_\alpha(t_i)) \right],
\end{eqnarray}
where we have partitioned the interval $[0,t]$ into $0=t_0<t_1<...<t_M=t$, where $h=t_{i+1}-t_{i}$. In the limit we obtain,
\begin{equation}\label{eq:survival_riemann_stieltjes}
    \mathbb{P}[\mathcal{T}>t]_{x_0} =\left\langle \exp \left\{-p \int_0^t \delta(X(S_\alpha(t))) dS_\alpha(t) \right\} \right \rangle_{x_0}.
\end{equation}
Due to the independence between the subdiffusive dynamics and the reaction events, we find that Eq. (\ref{eq:fractional_fk_sol}) (for $U(x)=\delta(x)$), can be understood as \cite{grebenkov2020paradigm,bressloff2022diffusion}
\begin{equation}\label{eq:frac_fk_sol_prob}
    P_\alpha(x,p,t\vert x_0)dx=\mathbb{P}[X(S_\alpha(t)) \in [x,x+dx),t<\mathcal{T}]_{x_0}.
\end{equation}
This shows $P_\alpha(x,p,t\vert x_0)$ can be interpreted as the probability density of a subdiffusive particle starting at $x_0$ to be at a position $x$ whilst having not reacted at the origin. To link Eq. (\ref{eq:frac_fk_sol_prob}) to the joint probability density of the subordinated local time and position, $\rho_\alpha(x,l,t\vert x_0)$, we use the integral form of $P_\alpha(x,p,t\vert x_0)$ in Eq. (\ref{eq:fractional_fk_sol}) and with $\rho_\alpha(x,l,t\vert x_0)dx=\partial_l \mathbb{P}[X(S_\alpha(t))\in[x,x+dx),\ell(S_\alpha(t))<l]_{x_0}$ we integrate by parts to find 
\begin{eqnarray}\label{eq:laplace_local_time_prob}
     P_\alpha(x,p,t\vert x_0)dx=\int_0^\infty dl pe^{-pl} \mathbb{P}[X(S_\alpha(t))\in[x,x+dx),\ell(S_\alpha(t))<l]_{x_0}.
\end{eqnarray}
Inside the integral we have the probability density of an exponentially distributed random variable, $\hat{l}$, with mean $1/p$, i.e. $\mathbb{P}[\hat{l}\in[l,l+dl)]=pe^{-pl}dl$. So if we replace $pe^{-pl}$ with $\langle \delta(l-\hat{l})\rangle$, we obtain
\begin{eqnarray}\label{eq:prob_def_local}
    P_\alpha(x,p,t\vert x_0)dx    =\mathbb{P}[X(S_\alpha(t)) \in [x,x+dx),\ell(S_\alpha(t))<\hat{l}]_{x_0}.
\end{eqnarray}
Therefore, we have shown that the solution of the generalized FKE (\ref{eq:fractional_fk}), $P_\alpha(x,p,t\vert x_0)$, is the probability density of a subdiffusing particle to be found at a position $x$, whilst the subordinated local time has not exceeded the value of an exponentially distributed random variable $\hat{l}$ (see discussion in Sec. \ref{sec:introduction} and Refs. \cite{grebenkov2019probability,grebenkov2020imperfect,grebenkov2020paradigm,bressloff2022diffusion}). As $\ell(S_\alpha(t))$ is a monotonically non-decreasing process, the event of $\{\hat{l}>\ell(S_\alpha(t))\}$ is equivalent to $\{\mathcal{T}>t\}$. Thus, by comparing Eq. (\ref{eq:prob_def_local}) to Eq. (\ref{eq:frac_fk_sol_prob}) we can see how the reaction time, $\mathcal{T}$, is related to the subordinated local time, $\ell(S_\alpha(t))$, via
\begin{equation}\label{eq:reaction_time_local}
    \mathcal{T}=\inf\{t>0:\ell(S_\alpha(t))>\hat{l} \}.
\end{equation}
In other words the reaction time and subordinated local time are intimately linked processes, with the reaction time being determined by the subordinated local time exceeding a certain value.

\subsection{Radiation Boundary Condition}\label{sec:rad_bc}

In the previous section we have established the connection between reactions and the generalized FKE, here we extend the idea to when the reactions are occurring on a boundary. We consider a CTRW generated by a nearest-neighbour random walk moving on a discrete one-dimensional lattice with sites $0,1,2,...$ and lattice spacing $\varepsilon$ with the waiting time distribution, $\psi(t)$, being heavy-tailed, thus $\psi(t)\simeq \upsilon^\alpha/t^{1+\alpha}$, where $\upsilon$ is a temporal scale parameter. The dynamics of the walker may be described by the GME for the occupation probability at lattice site $i$, initially starting from site $j$, $\mathcal{W}_{i,j}(t)$, for $i>0$ the GME is \cite{kenkre1973generalized}
\begin{equation}\label{eq:master_eq}
    \frac{d \mathcal{W}_{i,j}(t)}{dt}=\int_0^t dt' \Phi(t-t') \frac{\mathcal{W}_{i+1,j}(t')+\mathcal{W}_{i-1,j}(t')-2\mathcal{W}_{i,j}(t')}{2},
\end{equation}
where the Laplace transform of the memory kernel being, $\widetilde{\Phi}(\epsilon)=\epsilon \widetilde{\psi}(\epsilon)/(1-\widetilde{\psi}(\epsilon))$, here $\epsilon$ is again the Laplace variable. At the site $i=0$, we have
\begin{equation}\label{eq:master_eq_0}
    \frac{d \mathcal{W}_{0,j}(t)}{dt}=\int_0^t dt' \Phi(t-t') \frac{\mathcal{W}_{1,j}(t')-\mathcal{W}_{0,j}(t')}{2}.
\end{equation}

In the presence of a reactive boundary an incident particle at the lattice site $i=0$, has a probability of reacting $1-k$ and a probability of being reflected $k$. This situation can be summarized by the following flux condition,
\begin{equation}\label{eq:flux_bc}
    \mathcal{J}_{0,j}^+(t)=k\mathcal{J}_{0,j}^-(t).
\end{equation}
From Eq. (\ref{eq:master_eq_0}) we can identify the discrete fluxes in and out of the boundary as $\mathcal{J}^+_{0,j}(t)=\int_0^tdt'\Phi(t-t') \mathcal{W}_{0,j}(t')/2$ and $\mathcal{J}^-_{0,j}(t)=\int_0^tdt'\Phi(t-t') \mathcal{W}_{1,j}(t')/2$, respectively \cite{lomholt2007subdiffusion}. If we insert these fluxes into Eq. (\ref{eq:flux_bc}), and using the following relation for the total flux, $\mathcal{J}_{0,j}(t)=\mathcal{J}^+_{0,j}(t)-\mathcal{J}^-_{0,j}(t)$, we obtain
\begin{equation}
    \mathcal{J}_{0,j}(t)=\frac{k-1}{2} \int_0^t dt'\Phi(t-t') \mathcal{W}_{1,j}(t'),
\end{equation}
and find the relation in terms of $\mathcal{W}_{0,j}(t)$, as
\begin{equation}\label{eq:discrete_rad_bc}
    \mathcal{J}_{0,j}(t)=\frac{k-1}{2k}\int_0^t dt'\Phi(t-t')\mathcal{W}_{0,j}(t').
\end{equation}

Let us now take the diffusive limit, which entails taking the limits $\varepsilon \to 0$ and $\upsilon^\alpha \to 0$ \cite{barkai2000continuous}. This corresponds to $\widetilde{\psi}(\epsilon)\sim 1-(\upsilon \epsilon)^\alpha$, so $\widetilde{\Phi}(\epsilon) \sim \upsilon^{-\alpha} \epsilon^{1-\alpha}$ with $\mathcal{W}_{i,j}(t)/\varepsilon \to W_\alpha(x,t\vert x_0)$ and $\mathcal{J}_{i,j}(t)\to J_\alpha(x,t\vert x_0)$, where $i\varepsilon \to x$ and $j\varepsilon\to x_0$. By taking the Laplace transform of Eq. (\ref{eq:discrete_rad_bc}) and inserting $\widetilde{\Phi}(\epsilon)$, we obtain
\begin{equation}\label{eq:discrete_rad_bc_limit}
    \widetilde{J}_\alpha(0,\epsilon\vert x_0)=-\lambda_\alpha \epsilon^{1-\alpha}\widetilde{W}_\alpha(0,\epsilon\vert x_0).
\end{equation}
where the flux is defined as $J_\alpha(x,t\vert x_0)=-K_\alpha \, {}_0D_t^{1-\alpha} \partial_x W_\alpha(x,t\vert x_0)$ (from writing Eq. (\ref{eq:fractional_diffusion}) as a continuity equation, {}$\partial_t W_\alpha(x,t)+\partial_x J_\alpha(x,t)=0$) and 
\begin{equation}\label{eq:lambda_limit}
    \frac{1-k}{k \varepsilon} K_\alpha \to \lambda_\alpha,
\end{equation}
with $\varepsilon^2/(2 \upsilon^\alpha)\to K_\alpha$. Eq. (\ref{eq:lambda_limit}) implies that in the diffusive limit one requires $k\to 1$ \cite{grebenkov2003spectral}, which is a consequence of the number of visits to the origin becoming infinite. 

Taking the inverse Laplace transform of Eq. (\ref{eq:discrete_rad_bc_limit}), we have
\begin{equation}\label{eq:fractiona_rad_bc}
    J_\alpha(0,t\vert x_0)=-\lambda_\alpha \,{}_0D_t^{1-\alpha} W_\alpha(0,t\vert x_0).
\end{equation}
Clearly Eq. (\ref{eq:fractiona_rad_bc}) is equivalent to the so-called radiation BC \cite{weiss1994aspects,redner2001guide,grebenkov2010subdiffusion}, 
\begin{equation}\label{eq:rad_bc}
    K_\alpha \frac{\partial W_\alpha(x,t\vert x_0)}{\partial x}\Big\vert_{x=0}=\lambda_\alpha W_\alpha(0,t\vert x_0). 
\end{equation}
The radiation BC (\ref{eq:rad_bc}) describes a reactive boundary such that an incident particle is either absorbed or reflected depending on the reactivity parameter $\lambda_\alpha$ which has dimensions $[\text{length}]/[\text{time}]^\alpha$, such that $\lambda_\alpha=\infty$ represents full absorption and $\lambda_\alpha=0$ represents full reflection. 

Note that here we have obtained the reactivity parameter, $\lambda_\alpha$, derived from a reaction probability principle, whereas previously it had been found using a reaction rate \cite{seki2003recombination,seki2003fractional,lomholt2007subdiffusion,eaves2008subdiffusive}.

\subsection{Connection Between the Generalized Feynman-Kac Equation and the Radiation Boundary Condition}\label{sec:connection_FKE_rad}

Starting from the limit form of $\lambda_\alpha$ in Eq. (\ref{eq:lambda_limit}), for small $\varepsilon$ the probability of reflection can be expressed as $k\approx 1/(1+\varepsilon \lambda_\alpha/K_\alpha)$. The probability of visiting the origin $n$ times without a reaction (being reflected) is simply $k^n$ since each interaction is independent, but with reference to Sec. \ref{sec:sub_local_time}, we know the particle visits the origin a random number of times, $\mathcal{N}(t)$. To find the probability of the particle, which started at $x_0$, to have not reacted, we must average over all possible realisations of $\mathcal{N}(t)$, to which we obtain \cite{grebenkov2022encounter}
\begin{equation}\label{eq:ensemble_av_react}
    \mathbb{P}[\mathcal{T}>t]_{x_0}=\left\langle k^{\mathcal{N}(t)}\right\rangle_{x_0}.
\end{equation}
After inserting the expression for $k$ into Eq. (\ref{eq:ensemble_av_react}) and expanding for small $\varepsilon$, we have
\begin{equation}\label{eq:survival_approx}
    \mathbb{P}[\mathcal{T}>t]_{x_0} \approx \left\langle \exp\left\{-\frac{\varepsilon \mathcal{N}(t)\lambda_\alpha}{K_\alpha}\right\} \right\rangle_{x_0}.
\end{equation}
Using Eq. (\ref{eq:visits_macro}), Eq. (\ref{eq:survival_approx}) in the limit $\varepsilon\to 0$ becomes \cite{grebenkov2022encounter},
\begin{equation}\label{eq:survival_lim}
    \mathbb{P}[\mathcal{T}>t]_{x_0} = \left\langle e^{-\lambda_\alpha \ell(S_\alpha(t))}  \right\rangle_{x_0},
\end{equation}
which is exactly Eq. (\ref{eq:survival_riemann_stieltjes}) with $\lambda_\alpha$ in place of $p$. Note that because of the boundary $X(\tau)$ is now reflected Brownian motion, whereas in Sec. \ref{sec:frac_fk_react} we did not impose such a restriction. 

This implies that the solution to the FDE, (\ref{eq:fractional_diffusion}) with the radiation BC (\ref{eq:rad_bc}) is equivalent to solving the generalized FKE (\ref{eq:fractional_fk}) with $U(x)=\delta(x)$ and the reflecting (Neumann) BC, $\lim_{x\to 0} \partial_x P_\alpha(x,p,t\vert x_0)=0$. We can simply verify this by considering the generalized FKE with $U(x)=\delta(x)$,
\begin{equation}\label{eq:fractional_fk_delta}
    \frac{\partial}{\partial t}P_\alpha(x,\lambda_\alpha,t\vert x_0)={}_0D_t^{1-\alpha}\Big[K_\alpha \frac{\partial^2}{\partial x^2}    - \lambda_\alpha \delta(x)  \Big]P_\alpha(x,\lambda_\alpha,t\vert x_0),
\end{equation}
where we have replaced $p$ with $\lambda_\alpha$ as we are looking specifically at the radiation BC. Eq. (\ref{eq:fractional_fk_delta}) has been considered before in the specific context of geminate recombination \cite{seki2003fractional,seki2003recombination,seki2007specific}. Integrating both sides of Eq. (\ref{eq:fractional_fk_delta}) over the range $[0,\Delta]$ with respect to $x$ and making use of the reflecting BC, gives \cite{seki2003fractional}
\begin{eqnarray}\label{eq:radition_deriv}
     \int_0^\Delta \frac{\partial}{\partial t}P_\alpha(x,\lambda_\alpha,t\vert x_0) dx= {}_0D_t^{1-\alpha} \Big[K_\alpha\frac{\partial P_\alpha(x,\lambda_\alpha,t\vert x_0)}{\partial x} \Big\vert_{x=\Delta}\\ \nonumber
     -\lambda_\alpha  P_\alpha(0,\lambda_\alpha,t\vert x_0)\Big].
\end{eqnarray}
By taking $\Delta\to0$ the left-hand side vanishes and the radiation BC (\ref{eq:rad_bc}) is recovered, showing the equivalent formulations of the problem. We note that it has recently been shown that another equation (not in the FKE form), coming from (normal) diffusion through permeable barriers, satisfies the radiation boundary condition \cite{kay2022diffusion} with certain conditions. We extend this equation to the subdiffusive case in Appendix \ref{sec:permeable} and show how it leads to the radiation BC. This fact further compounds the notion that the reaction dynamics is a subordinated process.

\section{First-Reaction Time}\label{sec:first_reac_time}

\subsection{Generalized Feynman-Kac Equation Solution}

To study the first-reaction time (FRT) of a subdiffusive particle in the presence of a radiation boundary at the origin, we must find the corresponding probability density, $P_\alpha(x,\lambda_\alpha,t \vert x_0)$. As demonstrated in Sec. \ref{sec:connection_FKE_rad} this may be achieved through two equivalent methods, solving the FDE (\ref{eq:fractional_diffusion}) with the radiation BC (\ref{eq:rad_bc}) or solving the generalized FKE (\ref{eq:fractional_fk_delta}) with a reflecting BC at the origin. One can appreciate that the latter task is more amenable. Making use of the method of images \cite{montroll1979enriched} the solution simply follows through knowledge of the Green's function of Eq. (\ref{eq:fractional_fk_delta}), that is the solution of the FDE with a reflecting BC, which is given in the Laplace domain by \cite{metzler2000random}
\begin{equation}\label{eq:fde_reflect_prop}
    \widetilde{W}_\alpha(x,\epsilon\vert x_0)=\frac{e^{ -\vert x-x_0\vert\sqrt{\epsilon^\alpha/K_\alpha}}+e^{ -\vert x+x_0\vert\sqrt{\epsilon^\alpha/K_\alpha}}}{2\sqrt{K_\alpha \epsilon^{2-\alpha}}}.
\end{equation}
$P_\alpha(x,\lambda_\alpha,t\vert x_0)$, is then constructed in terms of the Green's function as,
\begin{multline}\label{eq:defect_integral}
    P_\alpha(x,\lambda_\alpha,t\vert x_0)=W_\alpha(x,t\vert x_0) \\ 
    -\lambda_\alpha  \int_{0}^{t} dt' \int_{0}^{\infty} dy W_\alpha(x,t-t'\vert y) \delta(y) {}_0D_{t'}^{1-\alpha}P_\alpha(y,\lambda_\alpha,t'\vert x_0).
\end{multline}
From the defect technique \cite{kenkre2021memory} Eq. (\ref{eq:defect_integral}) can be written in the Laplace domain as follows \cite{szabo1984localized},
\begin{eqnarray}\label{eq:frac_fk_sol}
    \widetilde{P}_\alpha(x,\lambda_\alpha,\epsilon\vert x_0)= \widetilde{W}_\alpha(x,\epsilon\vert x_0)    - \widetilde{W}_\alpha(x,\epsilon\vert0)\frac{ \widetilde{W}_\alpha(0,\epsilon\vert x_0)}{ \frac{1}{\lambda_\alpha \epsilon^{1-\alpha}}+\widetilde{W}_\alpha(0,\epsilon\vert0)}.
\end{eqnarray}
Substitution of Eq. (\ref{eq:fde_reflect_prop}) into Eq. (\ref{eq:frac_fk_sol}) gives,
\begin{equation}\label{eq:frac_fk_sol_laplace}
    \widetilde{P}(x,\lambda_\alpha,\epsilon\vert x_0)=\widetilde{W}_\alpha(x,\epsilon\vert x_0)-\frac{\lambda_\alpha  e^{-(x+x_0) \sqrt{\epsilon ^\alpha /K_\alpha}}}{\lambda_\alpha  \sqrt{K_\alpha \epsilon ^{2-\alpha }}+K_\alpha \epsilon },
\end{equation}
which can be readily shown to satisfy the radiation BC at $x=0$. 

\subsection{First-Reaction Time Probability Density}

To find the FRT probability density we first consider the survival probability Eq. (\ref{eq:survival_lim}), i.e. $Q_\alpha(t\vert x_0)=\mathbb{P}[\mathcal{T}>t]_{x_0}$, to which we may write in terms of the survival probability of normal diffusion, $Q(t\vert x_0)=\left\langle \exp\{-\lambda \int_0^t \delta(X(t'))dt'\}\right \rangle_{x_0}$, \cite{dienst2007mean}
\begin{equation}\label{eq:surival_subord}
    Q_\alpha(t\vert x_0)=\int_0^\infty g(\tau,t) \left\langle \exp\left\{-\lambda \int_0^\tau \delta(X(t'))dt'\right\}\right \rangle_{x_0} d\tau.
\end{equation}
As the FRT probability density, $\mathcal{F}_\alpha(t\vert x_0)$, is the density of the reaction times $\mathcal{T}$, it is then related to the survival probability via $\mathcal{F}_\alpha(t\vert x_0)=-\partial_t Q_\alpha(t\vert x_0)$. After using Eq. (\ref{eq:frac_diff_eq}) we find
\begin{equation}\label{eq:fpt_norm_sub}
    \mathcal{F}_\alpha(t\vert x_0)={}_0D_t^{1-\alpha} \int_0^\infty \left\langle \exp\left\{-\lambda \int_0^\tau \delta(X(t'))dt'\right\}\right \rangle_{x_0} \frac{\partial}{\partial \tau}g(\tau,t) d\tau,
\end{equation}
which can be integrated by parts leading to
\begin{equation}\label{eq:fpt_prob_rep}
    \mathcal{F}_\alpha(t\vert x_0)=\lambda_\alpha {}_0D_t^{1-\alpha} \left\langle \delta(X(S_\alpha(t))) e^{-\lambda_\alpha \int_0^{S_\alpha(t)} \delta(X(t'))dt'} \right\rangle_{x_0},
\end{equation}
where we have used $g(\tau,t)=\langle \delta(\tau-S_\alpha(t))\rangle$. Comparing Eq. (\ref{eq:fpt_prob_rep}) to Eq. (\ref{eq:fractional_fk_sol}) one can see that the FRT probability density is simply related to $P_\alpha(x,\lambda_\alpha,t\vert x_0)$, via
\begin{equation}\label{eq:fpt_fk_sol}
    \mathcal{F}_\alpha(t\vert x_0)=\lambda_\alpha {}_0D_t^{1-\alpha} P_\alpha(0,\lambda_\alpha,t\vert x_0).
\end{equation}
This relation is obvious from the radiation BC since the FRT probability density is equal to the (negative) flux at the boundary. Then from Eq. (\ref{eq:frac_fk_sol_laplace}) we find the Laplace Transform of the FRT density as \cite{eaves2008subdiffusive},
\begin{equation}\label{eq:laplace_fpt}
    \widetilde{\mathcal{F}}_\alpha(\epsilon\vert x_0)=\frac{\lambda_\alpha  e^{-x_0 \sqrt{\epsilon ^\alpha /K_\alpha}}}{\sqrt{K_\alpha \epsilon ^\alpha}+\lambda_\alpha}.
\end{equation}

Since Eq. (\ref{eq:laplace_fpt}) is hard to invert directly, we proceed by finding the Mellin transform, $\mathcal{M}\{ f(t)\}=\widehat{f}(s)=\int_0^\infty t^{s-1} f(t)dt$, and then performing the inverse transform \cite{wyss1986fractional,schneider1989fractional} (see Appendix \ref{sec:appen_mellin} and \ref{sec:appen_inverse_mellin}). In the end we obtain,
\begin{multline}\label{eq:fpt_time_sum}
     \mathcal{F}_\alpha(t\vert x_0)=\frac{e^{\lambda_\alpha x_0/K_\alpha}}{t} \Bigg[\frac{\lambda_\alpha t^{\alpha/2}}{\sqrt{K_\alpha}} E_{\frac{\alpha}{2},\frac{\alpha}{2}}\left(-\frac{\lambda_\alpha t^{\alpha/2}}{\sqrt{K_\alpha}}\right)\\ 
     - \sum_{n=0}^\infty \frac{(-1)^n \gamma\left(n+1,\frac{\lambda_\alpha x_0}{K_\alpha}\right)}{n! \Gamma\left(\frac{-\alpha n}{2}\right) \left(\frac{\lambda_\alpha t^{\alpha/2}}{\sqrt{K_\alpha}}  \right)^n} \Bigg],
\end{multline}
where $\Gamma(z)=\int_0^\infty t^{z-1} e^{-t}dt$ is the Gamma function \cite{abramowitz1964handbook}, $\gamma(z,a)=\int_0^a t^{z-1} e^{-t}dt$ is the lower incomplete Gamma function \cite{abramowitz1964handbook} and $E_{a,b}(z)$ is the two-parameter Mittag-Leffler function \cite{bateman1953higher3} (see Eq. (\ref{eq:mittag_leffler_def})). Using the integral form of the incomplete Gamma function, we may write Eq. (\ref{eq:fpt_time_sum}) in a more compact integral form,
\begin{multline}\label{eq:fpt_time_integral}
    \mathcal{F}_\alpha(t\vert x_0)=\frac{e^{\lambda_\alpha x_0/K_\alpha}}{t}\Bigg[\frac{\lambda_\alpha t^{\alpha/2}}{\sqrt{K_\alpha}} E_{\frac{\alpha}{2},\frac{\alpha}{2}}\left(-\frac{\lambda_\alpha t^{\alpha/2}}{\sqrt{K_\alpha}}\right)\\ 
     - \int_0^{\frac{\lambda_\alpha x_0}{K_\alpha}} e^{-x} \phi\left(-\frac{\alpha}{2},0;\frac{-\sqrt{K_\alpha}}{\lambda_\alpha t^{\alpha/2}}x\right) dx\Bigg],
\end{multline}
where $\phi(a,b;z)$ is the Wright function as defined by the series \cite{wright1935asymptotic,wright1940generalized}
\begin{equation}\label{eq:wright_function}
    \phi(a,b;z)=\sum_{n=0}^\infty \frac{z^n}{n! \Gamma(an+b)},
\end{equation}
for $a>-1$. From the Eqs. (\ref{eq:fpt_time_sum}) and (\ref{eq:fpt_time_integral}) we can see that if the particle initially starts at the origin we obtain the FRT density in the simple form,
\begin{equation}\label{eq:fpt_origin}
    \mathcal{F}_\alpha(t\vert0)= \frac{\lambda_\alpha }{\sqrt{K_\alpha}t^{1-\frac{\alpha}{2}}} E_{\frac{\alpha}{2},\frac{\alpha}{2}}\left(-\frac{\lambda_\alpha t^{\alpha/2}}{\sqrt{K_\alpha}}\right).
\end{equation}
Using the following Laplace transform relationship between the Wright function and the Mittag-Leffler function \cite{gorenflo1999analytical,luchko2019wright}, 
\begin{equation}\label{eq:wright_laplace}
    E_{a,b+a}(-z)=\int_0^\infty e^{-zx}\phi(-a,b,-x)dx
\end{equation}
for $a>0$, then Eq. (\ref{eq:fpt_time_integral}) becomes,
\begin{equation}\label{eq:fpt_integral}
    \mathcal{F}_\alpha(t\vert x_0)=\frac{e^{\lambda_\alpha x_0/K_\alpha}}{t} \int_{\frac{\lambda_\alpha x_0}{K_\alpha}}^\infty e^{-x} \phi\left(-\frac{\alpha}{2},0;\frac{-\sqrt{K_\alpha}}{\lambda_\alpha t^{\alpha/2}}x\right) dx.
\end{equation}
This form of the FRT in Eq. (\ref{eq:fpt_integral}) is useful for verifying the perfectly reacting/absorbing case ($\lambda_\alpha\to \infty)$ and the normal diffusion case ($\alpha\to 1$) (see Appendix \ref{sec:appen_perfect_reactive} and \ref{sec:appen_diffusion_limit}). Eq. (\ref{eq:fpt_integral}) is now used to find the short and long time asymptotic form of the FRT. After a change of variable in the integral in Eq. (\ref{eq:fpt_integral}) we use the expression 
\begin{equation}\label{eq:fpt_integral_change}
    \mathcal{F}_\alpha(t\vert x_0)=\frac{\lambda_\alpha e^{\lambda_\alpha x_0/K_\alpha}}{\sqrt{K_\alpha}t^{1-\frac{\alpha}{2}}} \int_{\frac{x_0}{\sqrt{K_\alpha t^\alpha}}}^\infty e^{\frac{-\lambda_\alpha t^{\alpha/2} y}{\sqrt{K_\alpha}}} \phi\left(-\frac{\alpha}{2},0;-y\right) dy,
\end{equation}
to find the large $t$ dependence since lower limit of the integral tends to zero. Thus at long times, using Eq. (\ref{eq:wright_laplace}), we have the approximate form of $\mathcal{F}_\alpha(t\vert x_0)$,
\begin{equation}\label{eq:fpt_long_t_approx}
    \mathcal{F}_\alpha(t\vert x_0)\approx \frac{\lambda_\alpha e^{\lambda_\alpha x_0/K_\alpha}}{\sqrt{K_\alpha}t^{1-\frac{\alpha}{2}}} E_{\frac{\alpha}{2},\frac{\alpha}{2}}\left(-\frac{\lambda_\alpha t^{\alpha/2}}{\sqrt{K_\alpha}}\right).
\end{equation}
Using the asymptotic form of the Mittag-Leffler function \cite{bateman1953higher3} we have the following asymptotic dependence for $t\to \infty$,
\begin{equation}\label{eq:fpt_long_t_asymptotic}
    \mathcal{F}_\alpha(t\vert x_0) \sim \frac{\alpha \sqrt{K_\alpha} e^{\lambda_\alpha x_0/K_\alpha}}{2 \lambda_\alpha \Gamma(1-\alpha/2)}t^{-1-\alpha/2},
\end{equation}
which confirms the $t^{-1-\alpha/2}$ dependence \cite{eaves2008subdiffusive} as well as that it possesses the same asymptotic dependence as in the perfectly reacting case \cite{rangarajan2000anomalous}, leading to an infinite mean \cite{yuste2004comment}. 

\begin{figure}
    \centering
    \includegraphics[width=1\textwidth]{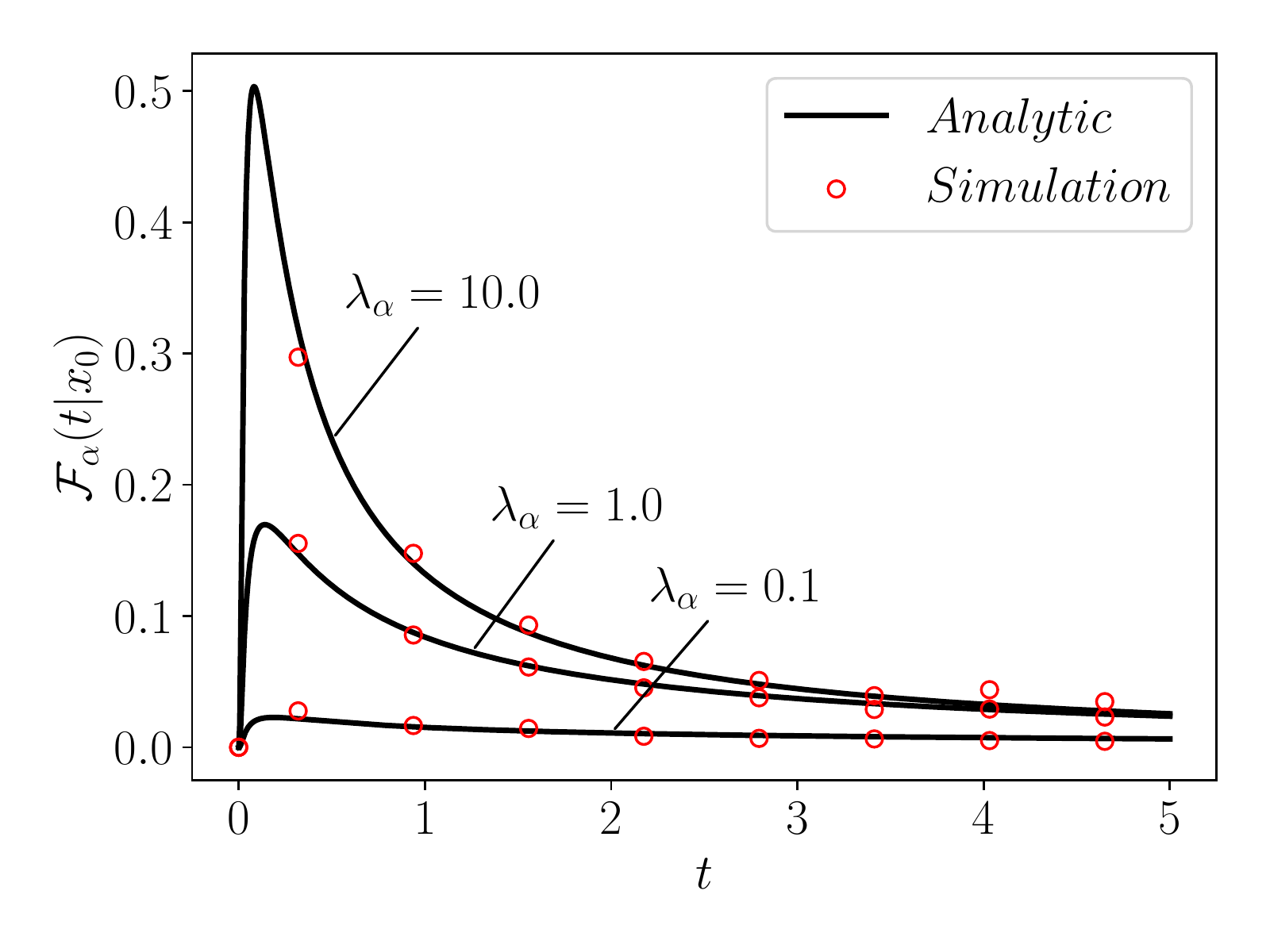}
    \caption{The first-reaction time density, $\mathcal{F}_\alpha(t\vert x_0)$, Eq. (\ref{eq:fpt_integral}) for $\alpha=2/3$, plotted for different values of $\lambda_\alpha$, with $x_0=2$ and $K_\alpha=2$, with all quantities in arbitrary units. We plot these curves against stochastic simulation results (shown as circles), as described in Sec. \ref{sec:simulations}.}
    \label{fig:fpt_sims}
\end{figure}

Similarly we utilise Eq. (\ref{eq:fpt_integral_change}) to study the short time asymptotics of $\mathcal{F}_\alpha(t\vert x_0)$, we use \cite{bateman1953higher3}
\begin{equation}\label{eq:wright_derivative}
    \partial_z \phi(a,b;z)=\phi(a,a+b;z)
\end{equation}
to integrate by parts giving and using $\lim_{z\to \infty}\phi\left(-\frac{\alpha}{2},\frac{\alpha}{2};-z\right)=0$ \cite{gorenflo1999analytical}, we find 
\begin{eqnarray}\label{eq:fpt_small_t_int}
   \mathcal{F}_\alpha(t\vert x_0)=\frac{\lambda_\alpha e^{\lambda_\alpha x_0/K_\alpha}}{\sqrt{K_\alpha}t^{1-\frac{\alpha}{2}}} \Bigg[e^{-\lambda_\alpha x_0/K_\alpha} \phi\left(-\frac{\alpha}{2},\frac{\alpha}{2};- \frac{x_0}{\sqrt{K_\alpha t^\alpha}} \right)  \\ \nonumber
    -\frac{\lambda_\alpha t^{\alpha/2}}{\sqrt{K_\alpha}} \int_{\frac{x_0}{\sqrt{K_\alpha t^\alpha}}}^\infty e^{\frac{-\lambda_\alpha t^{\alpha/2} y}{\sqrt{K_\alpha}}} \phi\left(-\frac{\alpha}{2},\frac{\alpha}{2};-y\right) dy  \Bigg].
\end{eqnarray}
For small $t$ we the can disregard the integral on the right-hand side (RHS) of Eq. (\ref{eq:fpt_small_t_int}), then using the asymptotic form of $\phi(a,b,-z)$ for $z\to \infty$ \cite{wright1940generalized,paris2016asymptotic}, we find $\mathcal{F}_\alpha(t\vert x_0)$ has the following dependence for $t\to 0$,
\begin{equation}\label{eq:fpt_small_t_asymp}
    \mathcal{F}_\alpha(t\vert x_0)\sim \frac{\alpha x_0 \lambda_\alpha \left(\frac{\alpha x_0 }{2 \sqrt{K_\alpha t^\alpha}}\right)^{\frac{1}{\alpha -2}}} {2\sqrt{(2-\alpha )\pi} K_\alpha t}\exp \left\{\frac{(\alpha -2)}{2} \left(\frac{\alpha}{2}\right)^{\frac{\alpha }{2-\alpha }} \left(\frac{x_0 }{\sqrt{K_\alpha t^\alpha}}\right)^{\frac{2}{2-\alpha}}\right\}.
\end{equation}
Eq. (\ref{eq:fpt_small_t_asymp}) indicates a short time exponential form for the FRT consistent with the perfectly reacting case \cite{rangarajan2000anomalous}. We plot $\mathcal{F}_\alpha(t\vert x_0)$ in Fig. (\ref{fig:fpt_sims}) to prove the validity of our analytic results. 

\subsection{Survival Probability and Subordinated Local Time Density}

Due to the reaction times, $\mathcal{T}$, being governed by the subordinated local time, $\ell(S_\alpha(t))$, we can find the distribution of $\ell(S_\alpha(t))$, $\rho_\alpha(l,t\vert x_0)$, and compare to simulations. From Eq. (\ref{eq:survival_lim}) we know that $\rho_\alpha(l,t\vert x_0)$ is the inverse Laplace transform of the survival probability $Q_\alpha(t\vert x_0)$ with respect to $\lambda_\alpha$. By finding the survival probability we thus have an easy route to determine the subordinated local time density. We find $Q_\alpha(t\vert x_0)=\int_t^\infty \mathcal{F}_\alpha(u\vert x_0)du$ by using Eq. (\ref{eq:fpt_integral}) and Eq. (\ref{eq:wright_derivative}) along with the fact that \cite{bateman1953higher3}
\begin{equation}\label{eq:wright_recurrence}
    az\phi(a,a+b;z)=\phi(a,b-1;z)+(1-b)\phi(a,b;z),
\end{equation}
to realize that $t^{-1}\phi\left(-\frac{\alpha}{2},0;\frac{-\sqrt{K_\alpha}}{\lambda_\alpha t^{\alpha/2}}x\right)=\frac{\partial}{\partial t} \phi\left(-\frac{\alpha}{2},1;\frac{-\sqrt{K_\alpha}}{\lambda_\alpha t^{\alpha/2}}x\right)$, which gives
\begin{equation}\label{eq:survival_exact}
    Q_\alpha(t\vert x_0)=1-e^{\lambda_\alpha x_0/K_\alpha} \int_{\frac{\lambda_\alpha x_0}{K_\alpha}}^\infty e^{-x} \phi\left(-\frac{\alpha}{2},1;\frac{-\sqrt{K_\alpha}}{\lambda_\alpha t^{\alpha/2}}x\right) dx.
\end{equation}
From Eq. (\ref{eq:survival_exact}) we recover the known solutions for the perfectly reactive and normal diffusion cases (see Appendix \ref{sec:appen_perfect_reactive} and \ref{sec:appen_diffusion_limit}).

Now we perfrom a change of variable on Eq. (\ref{eq:survival_exact}), which gives
\begin{equation}\label{eq:survival_exact_change}
    Q_\alpha(t\vert x_0)=1- \int_0^\infty \lambda_\alpha e^{-\lambda_\alpha l} \phi\left(-\frac{\alpha}{2},1;-  \frac{\sqrt{K_\alpha}}{t^{\alpha/2}}(l+x_0/K_\alpha)\right)dl,
\end{equation}
after integrating by parts and making use of Eq. (\ref{eq:wright_derivative}) this then leads to
\begin{multline}\label{eq:survival_parts}
     Q_\alpha(t\vert x_0)=1- \phi\left(-\frac{\alpha}{2},1;-  \frac{x_0}{\sqrt{K_\alpha t^\alpha}}\right) \\ 
    + \frac{\sqrt{K_\alpha}}{t^{\alpha/2}} \int_0^\infty  e^{-\lambda_\alpha l} \phi\left(-\frac{\alpha}{2},1-\frac{\alpha}{2};-  \frac{\sqrt{K_\alpha}}{t^{\alpha/2}}(l+x_0/K_\alpha)\right)dl.
\end{multline}
From Eq. (\ref{eq:survival_parts}) the inverse Laplace transform is straightforward and we find the subordinated local time density to be,
\begin{multline}\label{eq:local_time_exact}
     \rho_\alpha(l,t\vert x_0)=\delta(l)\left[1-  \phi\left(-\frac{\alpha}{2},1;-  \frac{x_0}{\sqrt{K_\alpha t^\alpha}}\right)\right]\\ 
    + \frac{\sqrt{K_\alpha}}{t^{\alpha/2}}\phi\left(-\frac{\alpha}{2},1-\frac{\alpha}{2};-\frac{\sqrt{K_\alpha}}{t^{\alpha/2}}(l+x_0/K_\alpha)\right).
\end{multline}
Clearly for a particle starting at the origin we obtain a simpler expression due to the first term on the RHS of Eq. (\ref{eq:local_time_exact}) vanishing. 

We are able to find the moments of $\rho_\alpha(l,t\vert x_0)$, $\langle \ell^n(S_\alpha(t)) \rangle_{x_0}=\int_0^\infty l^n \rho_\alpha(l,t\vert x_0) dl$, from the representation of $Q_\alpha(t\vert x_0)$ in Eq. (\ref{eq:survival_lim}), where $Q_\alpha(t\vert x_0)$ can be expressed as the moment generating function of $\ell(S_\alpha(t))$, such that
\begin{equation}\label{eq:mgf_survival}
    Q_\alpha(t\vert x_0)=\sum_{n=0}^{\infty}\frac{(-1)^n}{n!} \lambda_\alpha^n \langle \ell^n(S_\alpha(t)) \rangle_{x_0}.
\end{equation}
By repeatedly integrating by parts Eq. (\ref{eq:survival_exact}) whilst using Eq. (\ref{eq:wright_derivative}), we are able to express $Q_\alpha(t\vert x_0)$ as the following infinite series,
\begin{equation}\label{eq:survival_series}
    Q_\alpha(t\vert x_0)=\sum_{n=0}^{\infty} (-1)^n \left(\frac{\lambda_\alpha t^{\alpha/2}}{\sqrt{K_\alpha}}\right)^n \phi \left(-\frac{\alpha}{2},1+ \frac{\alpha}{2}n; \frac{-x_0}{\sqrt{K_\alpha t^\alpha}}\right).
\end{equation}
Thus we find the moments are uniquely given in terms of Wright functions,
\begin{equation}\label{eq:lt_moments}
    \langle \ell^n(S_\alpha(t)) \rangle_{x_0}=n! \left(\sqrt{\frac{t^\alpha}{K_\alpha}} \right)^n\phi \left(-\frac{\alpha}{2},1+ \frac{\alpha}{2}n; \frac{-x_0}{\sqrt{K_\alpha t^\alpha}}\right).
\end{equation}
For $x_0=0$ we obtain a simple form for the moments,
\begin{equation}\label{eq:lt_moments_zero}
    \langle \ell^n(S_\alpha(t)) \rangle_{0}=\frac{n!}{\Gamma(1+\frac{\alpha}{2}n)} \left(\sqrt{\frac{t^\alpha}{K_\alpha}} \right)^n,
\end{equation}
which can be realised by using the series form of the Mittag-Leffler function in Eq. (\ref{eq:fpt_origin}).

Interestingly, for the specific value of $\alpha=2/3$ \cite{gorenflo1999analytical}, we find the subordinated local time density to be in terms of the Airy function, $\mbox{Ai}(z)=\frac{1}{\pi}\int_0^\infty \cos(\frac{t^3}{3}+zt) dt$  for $\Im(z)=0$ \cite{bateman1953higher2}, 
\begin{equation}\label{eq:lt_airy}
    \rho_{2/3}(l,t\vert0)=3^{2/3} \sqrt{\frac{K_{2/3}}{t^{2/3}}} \mbox{Ai}\left( \sqrt{\frac{K_{2/3}}{(3t)^{2/3}}} l\right).
\end{equation}
For $\alpha=1$, one can see that Eq. (\ref{eq:local_time_exact}) reduces to the well known Gaussian solution (see Appendix \ref{sec:appen_diffusion_limit}). In Figure (\ref{fig:subordinated_local_time}), Eq. (\ref{eq:lt_airy}) is utilised to show the correct method of simulating the subordinated local time, as described in the next section. 

\begin{figure}
    \centering
    \includegraphics[width=1\textwidth]{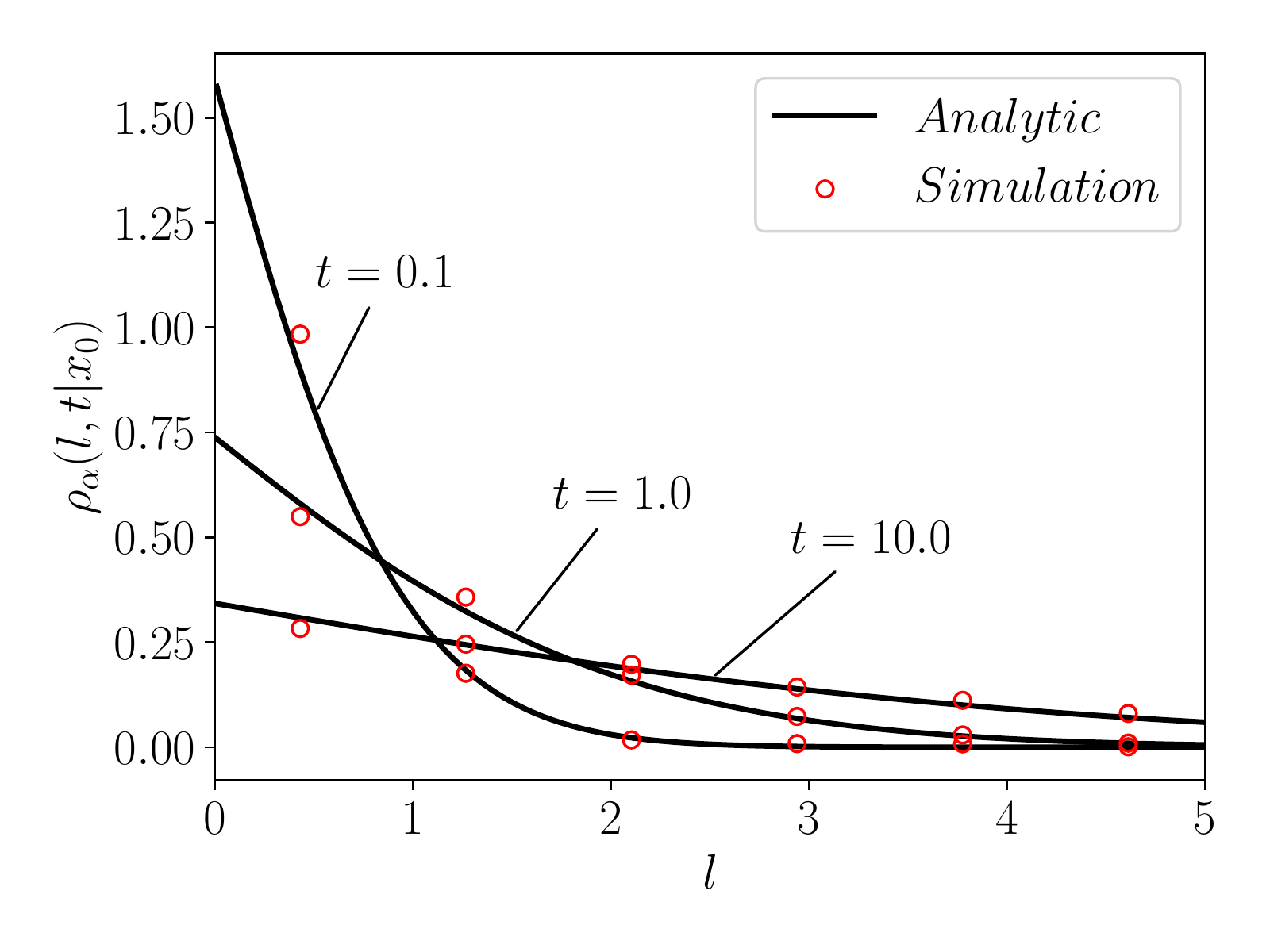}
    \caption{The subordinated local time density, $\mathcal{\rho}_\alpha(l,t\vert x_0)$, Eq. (\ref{eq:lt_airy}) for $\alpha=2/3$, plotted for different values of $t$, with $x_0=0$ and $K_\alpha=1$, with all quantities in arbitrary units. We plot these lines against stochastic simulations (shown as circles) of the subordinated local time, as described in Sec. \ref{sec:simulations}.}
    \label{fig:subordinated_local_time}
\end{figure}

\subsection{Simulations}\label{sec:simulations}

Due to the complexity of our analytic results it is important to validate with numerical simulations. A common method for simulating subdiffusion is through the CTRW formalism in the diffusive limit with a heavy-tailed waiting time distribution, however by keeping in the theme of this work we generate our simulations using a subordinated approach \cite{magdziarz2007fractional}, as described in Fig. (\ref{fig:subdiffusion_subordinate_sim}). Alternatively, one can simulate the subdiffusion trajectories via the algorithm in Ref. \cite{kleinhans2007continuous}, which differs from Ref. \cite{magdziarz2007fractional} through not needing to explicity calculate $S_\alpha(t)$. The aforementioned subordinated approach lends naturally to the simulation of the subordinated local time. We are able to effectively approximate, $\ell(S_\alpha(t))$ via the discrete construction in Sec. \ref{sec:sub_local_time}, where we approximate the limit in Eq. (\ref{eq:visits_macro}) for $\varepsilon<<1$. We approximate the integral in Eq. (\ref{eq:visits_lim}) by writing it in Riemann-Stieltjes form and using the left sided Riemann summation approximation, i.e. by discretizing the time interval, $[0,t]$, as $0=t_0<t_1<...<t_M=t$, then
\begin{equation}\label{eq:sub_local_time_approx}
    \ell(S_\alpha(t))\approx\frac{1}{\varepsilon} \sum_{i=0}^{M-1} \mathbb{I}_{\partial \Omega}(Y(t_i)) [S_\alpha(t_{i+1})-S_\alpha(t_i)].
\end{equation}
As we are considering a bounded domain, the subdiffusive trajectories are made to be reflected at the origin. As pointed out in Ref. \cite{grebenkov2019probability} one must ensure $\varepsilon>>\sqrt{2 K_\alpha \Delta t}$, where $\Delta t=t/M$, such that the characteristic size of the jump length is not larger than the region $\partial \Omega$, but the smaller the value of $\varepsilon$ the better the approximation of the subordinated local time. The simulation for the FRT density naturally follows from the subordinated local time, where for each trajectory we draw a random number from an exponential distribution with mean $1/\lambda_\alpha$ and record the FRT when the subordinated local time exceeds the value of the random number. Given the good match between simulations and theory in Figs. (\ref{fig:fpt_sims}) and (\ref{fig:subordinated_local_time}) this provides validation for the analytic results presented. 

\section{Summary and Conclusions}\label{sec:conclusion}

In summary we have derived, through subordination techniques, a type of generalized FKE and used such an equation to understand and analyse subdiffusion in the presence of a reactive boundary. After deriving the $\alpha$-stable inverse subordinator we perform a time change on the classical FKE to construct a generalized analogue, and find its solution. We introduce the notion of the subordinated local time and we interpret it as the continuum limit of the number of times a subdiffusive particle reaches a given location via a CTRW formulation. An important finding that emerges is that the time spent for each visit does not play a role in the subordinated local time. We apply the formalism to demonstrate how the generalized FKE can be used to describe a subdiffusion process in the presence of reactions. For that we show that the generalized FKE can be thought of as the position density of a subdiffusing particle with the requirement that the subordinated local time is less than a random variable drawn from an exponential distribution with a mean related to the parameter $p$ in the generalized FKE.

We also consider what would be the relevant BC if the reaction occurred on a boundary. We study this aspect by considering a CTRW on a discrete lattice and introduce a reflection probability at the origin, $k$, which gives a general flux condition. From this condition we take the relevant limits to obtain the generalized form of the radiation BC associated with subdiffusion and find the generalized reactivity parameter, $\lambda_\alpha$. We then demonstrate the equivalence between the generalized FKE with a reflecting boundary and the FDE with a radiation boundary. 

We employ the generalized FKE to study the FRT of a subdiffusive particle in the presence of a radiation boundary at the origin. We solve the generalized FKE using Green's function techniques and obtain an analytic solution in the Laplace domain. We then find the relation between the solution of the generalized FKE and the FRT probability density and obtain this quantity in the Laplace domain, which we are able to invert by converting it into a Mellin transform. The FRT probability density is obtained in terms of the Mittag-Leffler function and an infinite series of the lower incomplete Gamma functions or alternatively as an integral involving the Wright function. From this we are able to analyse the short and long time asymptotic form of this density recovering expected dependencies. Due to the fundamental connection between the FRT and subordinated local time, we calculate the subordinated local time density and all its moments. Finally, we show how our analytic results match with simulations, proving the validity of simulating subdiffusion in the presence of a radiation boundary using the subordinated local time approach. 

A natural extension to this work would be to consider the subordinated occupation time functional ($U(x)=\mathbb{I}_\Omega(x)$, for some spatial region $\Omega$) and how this may be used as in a similar sense as the subordinated local time here to describe subdiffusion with reactions in a certain region, not just at a boundary. Further future directions could include developing the backward version of the generalized FKE considered here to study the density of various other subordinated functionals and look at how they compare to functionals of subdiffusion \cite{carmi2010distributions}. Clearly functionals dependent on the underlying subdiffusive path like local time, occupation time etc. are certainly going to be different. However, the generalized FKE may be applicable to a class of functionals associated with first-passage times, due to only needing the knowledge of whether the particle has reached a specific point, not how long it has been there.

\bmhead{Acknowledgments}

We thank Eli Barkai for useful discussions. This work was carried out using the computational facilities of the Advanced Computing Research Centre, University of Bristol - http://www.bristol.ac.uk/acrc/. 

\section*{Declarations}

\bmhead{Funding}

TK and LG acknowledge funding from, respectively, an Engineering and Physical Sciences Research Council (EPSRC) DTP student grant and the Biotechnology and Biological Sciences Research Council (BBSRC) Grant No. BB/T012196/1

\bmhead{Data availability}

The datasets generated during and/or analysed during the current study are available from the corresponding author on reasonable request.

\bmhead{Conflicts of interest/Competing interests}

The authors have no conflicts of interest to declare that are relevant
to the content of this article.

\begin{appendices}

\section{Non Feynman-Kac Form of Equation Satisfying Radiation Boundary Condition}\label{sec:permeable}

We consider the following generalization of the so-called inhomogeneous diffusion equation introduced in Eq. (3) of Ref. \cite{kay2022diffusion} for subdiffusion,
\begin{eqnarray}\label{eq:fractional_permeable}
    \frac{\partial}{\partial t}P_\alpha(x,\lambda_\alpha,t\vert x_0)={}_0D_t^{1-\alpha}\Big[K_\alpha \frac{\partial^2}{\partial x^2}P_\alpha(x,\lambda_\alpha,t\vert x_0)   \\ \nonumber
     - \frac{K_\alpha^2}{\lambda_\alpha} \delta'(x)  \frac{\partial }{\partial x} P_\alpha(x,\lambda_\alpha,t\vert x_0) \big \vert_{x=0} \Big]
\end{eqnarray}
with the absorbing (Dirichlet) BC $P_\alpha(0,\lambda_\alpha,t\vert x_0)=0$. Let us integrate Eq. (\ref{eq:fractional_permeable}), as follows 
\begin{eqnarray}\label{eq:radiation_permeable_deriv}
    \int_{0}^{\Delta} dx \int dx \frac{\partial}{\partial t}P_\alpha(x,\lambda_\alpha,t\vert x_0)= K_\alpha \, {}_0D_t^{1-\alpha} P_\alpha(\Delta,\lambda_\alpha,t\vert x_0)\\ \nonumber
    - \frac{K_\alpha^2}{\lambda_\alpha} {}_0D_t^{1-\alpha} \frac{\partial }{\partial x} P_\alpha(x,\lambda_\alpha,t\vert x_0) \big \vert_{x=0}.
\end{eqnarray}
Then in the limit $\Delta\to 0$, Eq. (\ref{eq:radiation_permeable_deriv}) tends to the radiation BC, Eq. (\ref{eq:rad_bc}). Without the inclusion of the absorbing BC, Eq. (\ref{eq:fractional_permeable}) satisifes the permeable/leather BC \cite{tanner1978transient,powles1992exact}, where $\lambda_\alpha$ would be interpreted as a permeability.

\section{Mellin Transform of First-Reaction Time Density}\label{sec:appen_mellin}

Here we find the Mellin transform of Eq. (\ref{eq:laplace_fpt}), this is obtained by using the following relationship between the Laplace transform of a function and the Mellin transform \cite{schneider1989fractional}, where for an arbitrary function we have
\begin{equation}\label{eq:laplace_to_mellin}
    \widehat{f}(s)=\frac{1}{\Gamma(1-s)} \int_0^\infty \epsilon^{-s} \widetilde{f}(\epsilon) d\epsilon.
\end{equation}
Then from Eq. (\ref{eq:laplace_to_mellin}) with Eq. (\ref{eq:laplace_fpt}) the Mellin transform is found from the integral,
\begin{equation}\label{eq:fpt_mellin_integral}
    \widehat{\mathcal{F}}_\alpha(s\vert x_0)=\frac{1}{\Gamma(1-s)} \int_0^\infty \epsilon^{-s} \frac{B  e^{A \epsilon^{\alpha/2} }}{\epsilon^{\alpha/2}+B} d\epsilon,
\end{equation}
where for simplicity we have introduced the quantities $A=x_0/\sqrt{K_\alpha}$ and $B=\lambda_\alpha/\sqrt{K_\alpha}$. Now let us perform a change of variable for the integral in Eq. (\ref{eq:fpt_mellin_integral}) such that $u=A \epsilon^{\alpha/2}$, thus 
\begin{equation}\label{eq:fpt_mellin_integral_change}
    \widehat{\mathcal{F}}_\alpha(s\vert x_0)=\frac{2 B A^{\frac{2s-2}{\alpha}+1}}{\alpha \Gamma(1-s)} \int_0^\infty u^{\frac{2-2s}{\alpha}-1} \frac{  e^{-u} }{u+AB} du.
\end{equation}
Using a certain integral representation of the upper incomplete Gamma function \cite{bateman1953higher2}, $\Gamma(z,a)$,
\begin{equation}\label{eq:incomplete_gamma_integral}
    \Gamma(z,a)=\frac{a^z e^{-a}}{\Gamma(1-z)}\int_0^\infty \frac{e^{-x}}{x^z(x+a)} dx,
\end{equation}
we find the Mellin transform of the FRT density to be,
\begin{equation}\label{eq:mellin_transform_fpt}
    \widehat{\mathcal{F}}_\alpha(s\vert x_0)=\frac{2 e^{A B}B^{\frac{2-2s}{\alpha}} \Gamma\left( \frac{2-2s}{\alpha}\right) \Gamma\left( \frac{2s-2}{\alpha}-1,AB \right)}{\alpha\Gamma(1-s)}.
\end{equation}

\section{Inverse Mellin Transform of First-Reaction Time Density}\label{sec:appen_inverse_mellin}

\begin{figure}
    \centering
    \includegraphics[width=0.7\textwidth]{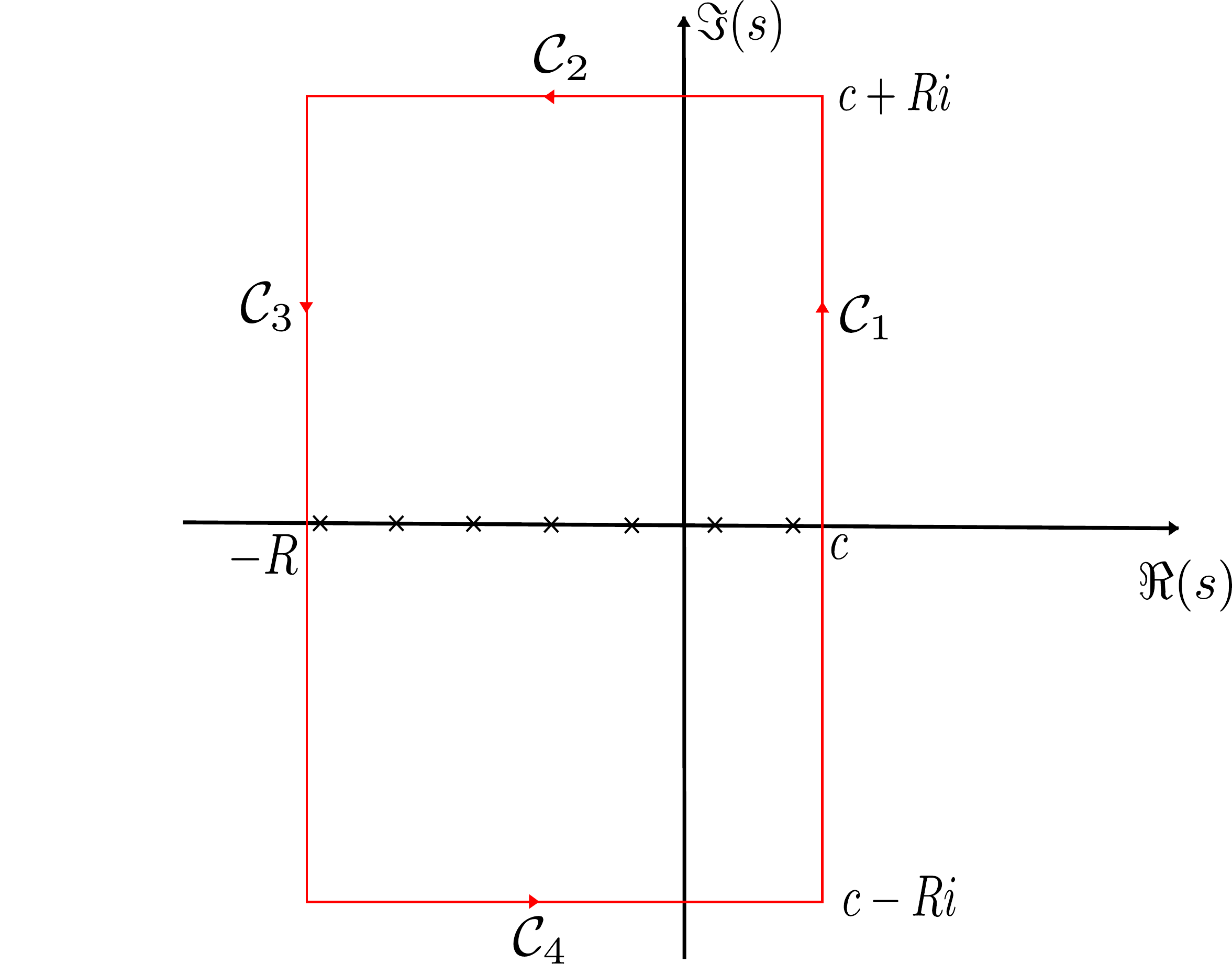}
    \caption{Contour, $\mathcal{C}=\mathcal{C}_1+\mathcal{C}_2+\mathcal{C}_3+\mathcal{C}_4$, for the inverse Mellin transform in Eq. (\ref{eq:contour_integral}). The crosses indicate poles.}
    \label{fig:Mellin_left_contour}
\end{figure}

Before performing the Mellin inverse transform of Eq. (\ref{eq:mellin_transform_fpt}) we write the upper incomplete Gamma function as $\Gamma(z,a)=\Gamma(z)-\gamma(z,a)$. So we perform a Mellin inversion on each part separately. Thus for the first part we have,
\begin{multline}\label{eq:contour_integral}
     \mathcal{M}^{-1}\left\{2 e^{AB}\frac{B^{\frac{2-2s}{\alpha}} \Gamma\left( \frac{2-2s}{\alpha}\right) \Gamma\left( \frac{2s-2}{\alpha}+1\right)}{\alpha\Gamma(1-s)} \right\} \\
    =\frac{2 B^{2/\alpha}e^{AB}}{\alpha 2 \pi i} \int_{c-i \infty}^{c+i\infty} \frac{ \Gamma\left( \frac{2-2s}{\alpha}\right) \Gamma\left( \frac{2s-2}{\alpha}+1\right)}{\Gamma(1-s)} (B^{\frac{2}{\alpha}} t)^{-s} ds,
\end{multline}
with $A$ and $B$ as defined in Appendix \ref{sec:appen_mellin}. Here $c$ is a real number lying in a strip where the Mellin transformed function is analytic and tends uniformly to zero as $\Im(s)\to \pm \infty$ \cite{kochubei2019handbook}. Due to the poles of $\Gamma(z)$ being located at $-n$ for $n=0,1,2,...$, then we can see the poles of the integrand of Eq. (\ref{eq:contour_integral}) are located at $s=1+n\alpha/2$ and $s=1-(n+1)\alpha/2$, so $1-\alpha/2<c<1+\alpha/2$. Due to the integrand diverging for $ s \to \infty$, to perform the integral in Eq. (\ref{eq:contour_integral}), we consider the contour $\mathcal{C}=\mathcal{C}_1+\mathcal{C}_2+\mathcal{C}_3+\mathcal{C}_4$ shown in Fig. (\ref{fig:Mellin_left_contour}). For $R\to \infty$, one can see that contributions from $\mathcal{C}_2,\mathcal{C}_3,\mathcal{C}_4$ vanish and thus computing the contour, $\mathcal{C}$, integral  is equivalent to performing the inverse Mellin transform in Eq. (\ref{eq:contour_integral}). Then from Cauchy's residue theorem, we only need to compute the residues at the poles, $s=1-(n+1)\alpha/2$.
Using $\mbox{Res}(\Gamma(z),-n)=(-1)^n/n!$, we then find the residues to be
\begin{eqnarray}
     \mbox{Res} \Bigg( \frac{ \Gamma\left( \frac{2-2s}{\alpha}\right) \Gamma\left( \frac{2s-2}{\alpha}+1\right)}{\Gamma(1-s)}(B^{2/\alpha} t)^{-s}, s= 1-\frac{(n+1)\alpha}{2}\Bigg) \\\nonumber
       =\frac{\alpha (-1)^n(B^{2/\alpha} t)^{\frac{\alpha+\alpha n}{2}-1}}{2 \Gamma\left(\frac{\alpha+\alpha n}{2} \right)},
\end{eqnarray}
leading to 
\begin{eqnarray}
     \mathcal{M}^{-1}\left\{2 e^{AB}\frac{B^{\frac{2-2s}{\alpha}} \Gamma\left( \frac{2-2s}{\alpha}\right) \Gamma\left( \frac{2s-2}{\alpha}+1\right)}{\alpha\Gamma(1-s)} \right\}
    =\frac{B e^{AB}}{t^{1-\alpha/2}} \sum_{n=0}^{\infty} \frac{(-B t^{\alpha/2})^n}{\Gamma\left( \frac{\alpha+\alpha n}{2}\right)}.
\end{eqnarray}
From the series representation of the Mittag-Leffler function \cite{bateman1953higher3},
\begin{equation}\label{eq:mittag_leffler_def}
    E_{a,b}(z)=\sum_{n=0}^{\infty} \frac{z^n}{\Gamma(an+b)},
\end{equation}
one obtains the first term on the RHS of Eq. (\ref{eq:fpt_time_sum}) in the main text. 

\begin{figure}
    \centering
    \includegraphics[width=0.7\textwidth]{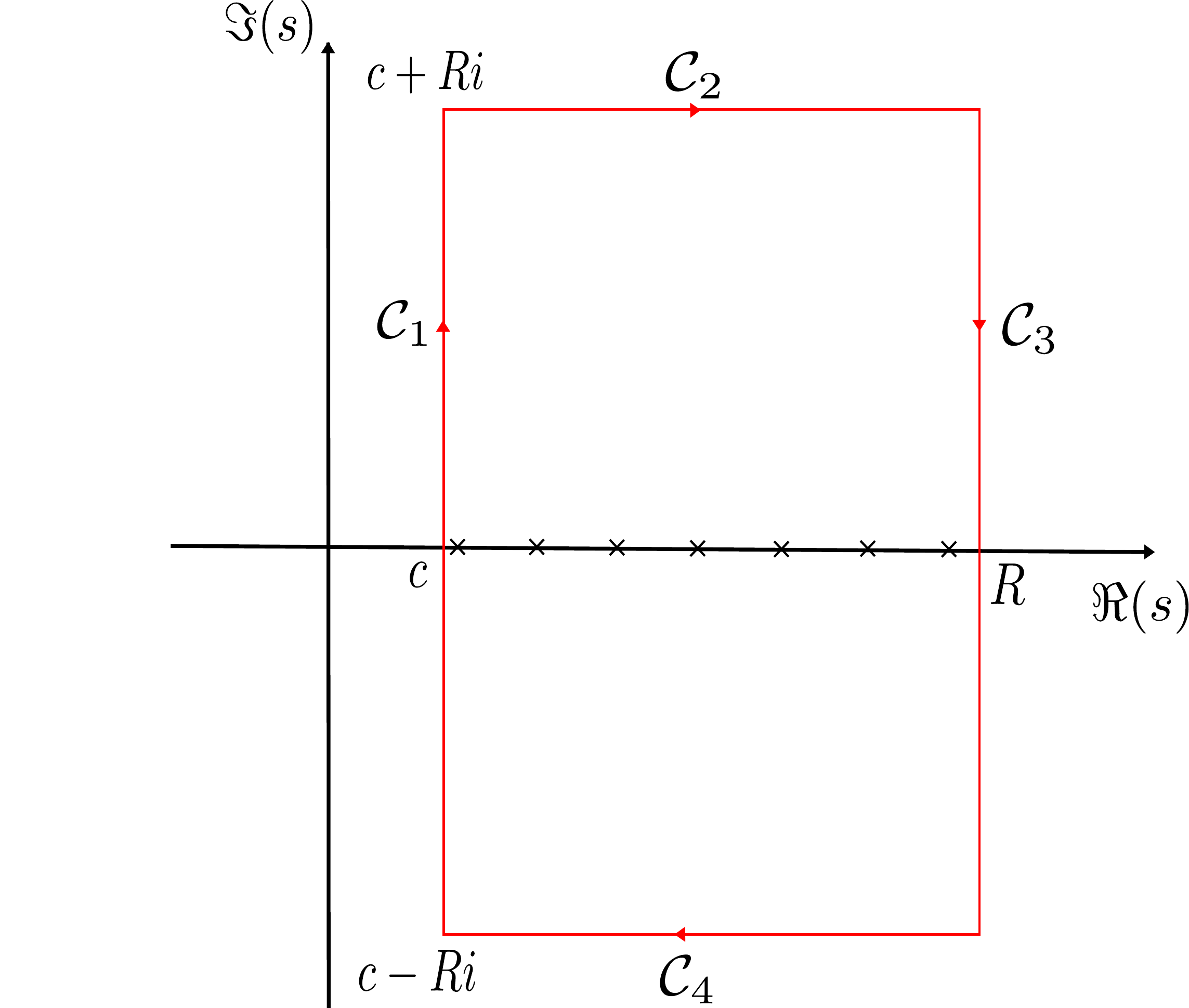}
    \caption{Contour, $\mathcal{C}=\mathcal{C}_1+\mathcal{C}_2+\mathcal{C}_3+\mathcal{C}_4$ for the inverse Mellin transform in Eq. (\ref{eq:inverse_mellin_right}). The crosses indicate poles.}
    \label{fig:Mellin_right_contour}
\end{figure}

Now for the other part of Eq. (\ref{eq:mellin_transform_fpt}) we first decompose the lower incomplete Gamma function into $\gamma(z,a)=a^z \Gamma(z) \gamma^*(z,a)$, where $\gamma^*(z,a)$ is an entire function \cite{abramowitz1964handbook}. Thus the inverse Mellin transform is computed via,
\begin{eqnarray}\label{eq:inverse_mellin_right}
    \mathcal{M}^{-1}\left\{2 e^{AB}\frac{B^{\frac{2-2s}{\alpha}} \Gamma\left( \frac{2-2s}{\alpha}\right) \gamma\left( \frac{2s-2}{\alpha}+1,AB\right)}{\alpha\Gamma(1-s)} \right\}
     =\frac{2 A^{1-2/\alpha}B e^{AB}}{\alpha 2 \pi i} \nonumber\\
     \times \int_{c-i \infty}^{c+i\infty} \frac{ \Gamma\left( \frac{2-2s}{\alpha}\right) \Gamma\left( \frac{2s-2}{\alpha}+1\right)\gamma^*\left( \frac{2s-2}{\alpha}+1,AB\right)}{\Gamma(1-s)} (A^{-\frac{2}{\alpha}} t)^{-s} ds,
\end{eqnarray}
so we can see the integrand has the same poles as previously and again $1-\alpha/2<c<1+\alpha/2$. However, due to the the integrand diverging for $s \to -\infty$ we perform the inverse Mellin transform in Eq. (\ref{eq:inverse_mellin_right}) using the contour $\mathcal{C}=\mathcal{C}_1+\mathcal{C}_2+\mathcal{C}_3+\mathcal{C}_4$ shown in Fig. (\ref{fig:Mellin_right_contour}). Again we take $R\to \infty$ and the contributions from $\mathcal{C}_2,\mathcal{C}_3,\mathcal{C}_4$ vanish, meaning the inverse Mellin transform is computed by finding the residues at the poles $s=1+n\alpha/2$. So the residues are, 
\begin{multline}
      \mbox{Res}\Bigg( \frac{ \Gamma\left( \frac{2-2s}{\alpha}\right) \Gamma\left( \frac{2s-2}{\alpha}+1\right) \gamma^*\left( \frac{2s-2}{\alpha}+1,AB\right)}{\Gamma(1-s)}(A^{-2/\alpha}t)^{-s},  s=1+\frac{n\alpha}{2}\Bigg)  \\
      =\frac{\alpha (-1)^{n+1}(A^{-2/\alpha} t)^{\frac{-\alpha n}{2}-1}}{2 \Gamma\left(\frac{-\alpha n}{2} \right)}\gamma^*(n+1,AB).
\end{multline}
Then we find the Mellin transform to be,
\begin{multline}
    \mathcal{M}^{-1}\left\{2 e^{AB}\frac{B^{\frac{2-2s}{\alpha}} \Gamma\left( \frac{2-2s}{\alpha}\right) \gamma\left( \frac{2s-2}{\alpha}+1,AB\right)}{\alpha\Gamma(1-s)} \right\}  \\
    =\frac{AB e^{AB}}{t} \sum_{n=0}^{\infty} \frac{(-1)^n(A t^{-\alpha/2})^n\gamma^*(n+1,AB)}{\Gamma\left(\frac{-\alpha n}{2} \right)},
\end{multline}
then after writing back in terms of $\gamma(z,a)$ we recover the summation in Eq. (\ref{eq:fpt_time_sum}). Note this sum should start from $n=1$ due $1/\Gamma(0)=0$ for $n=0$, but we leave it in this form for easier relation to the Wright function.

\section{Perfectly Reactive Case of First-Reaction Time Density and Survival Probability}\label{sec:appen_perfect_reactive}

We start from Eq. (\ref{eq:fpt_integral}) and make the change of variable, $y=x-AB$, with $A$ and $B$ as defined in Appendix \ref{sec:appen_mellin}, which gives
\begin{equation}\label{eq:fpt_change}
    \mathcal{F}_\alpha(t\vert x_0)=t^{-1} \int_{0}^\infty e^{-y} \phi\left(-\frac{\alpha}{2},0;-\frac{y}{B t^{\alpha/2}}-\frac{A}{t^{\alpha/2}}\right) dx.
\end{equation}
Now we take the perfectly reactive limit, $B\to\infty$, which after taking this inside the integral gives,
\begin{equation}\label{eq:fpt_perf_limit}
    \mathcal{F}_\alpha(t\vert x_0)=t^{-1} \phi \left(-\frac{\alpha}{2},0;-\frac{A}{t^{\alpha/2}}\right).
\end{equation}
Then by using Eq. (\ref{eq:wright_recurrence}), Eq. (\ref{eq:fpt_perf_limit}) becomes
\begin{equation}
    \mathcal{F}_\alpha(t\vert x_0)=\frac{\alpha A}{2 t^{1+\alpha/2}}\phi\left(-\frac{\alpha}{2},1-\frac{\alpha}{2};-\frac{A}{t^{\alpha/2}}\right),
\end{equation}
which is the exact solution found in \cite{rangarajan2000anomalous}. Using the relationship \cite{kochubei2019handbook} between the Wright function and the Fox H function \cite{fox1961G,mathai1978h,mathai2009h}, $\mathcal{F}_\alpha(t\vert x_0)$ is found in terms of the very general Fox H function \cite{rangarajan2000anomalous}
\begin{eqnarray}\label{eq:fpt_fox}
        \mathcal{F}_\alpha(t\vert x_0)=\frac{\alpha A}{2 t^{1+\alpha/2}}H_{1,1}^{1,0}\left(\frac{A}{t^{\alpha/2}}\Bigg\vert
    \begin{array}{c}
    (1-\alpha/2,\alpha/2) \\
    (0,1)
    \end{array}
    \right).
\end{eqnarray}

Performing the same change of variables for the survival probability, Eq. (\ref{eq:survival_exact}), and taking the limit $B\to \infty$ we have,
\begin{equation}\label{eq:survival_abs}
    Q_\alpha(t\vert x_0)=1-\phi \left(-\frac{\alpha}{2},1;-\frac{A}{t^{\alpha/2}}\right)
\end{equation}
which in terms of the Fox H function we have the known solution \cite{yuste2007subdiffusive},
\begin{eqnarray}\label{eq:survival_fox}
        Q_\alpha(t\vert x_0)=1-H_{1,1}^{1,0}\left(\frac{A}{t^{\alpha/2}}\Bigg\vert
    \begin{array}{c}
    (1,\alpha/2) \\
    (0,1)
    \end{array}
    \right).
\end{eqnarray}
Note, Eq. (\ref{eq:survival_abs}) appears in the expression for the subordinated local time density, showing the close relationship between the two quantities. 

\section{Normal Diffusion Case of First-Reaction Time Density, Survival Probability and Subordinated Local Time Density}\label{sec:appen_diffusion_limit}

For $\alpha= 1$ we have the following specific value of the Wright function \cite{gorenflo1999analytical},
\begin{equation}\label{eq:wright_spec}
    \phi\left(-\frac{1}{2},0;z\right)=-\frac{z}{2\sqrt{\pi}} e^{-z^2/4},
\end{equation}
then inserting into Eq. (\ref{eq:fpt_integral}) we have 
\begin{equation}\label{eq:erf_integration}
    \mathcal{F}(t\vert x_0)=\frac{e^{A B}}{B\sqrt{4 \pi t^3}} \int_{A B}^\infty e^{-x} x e^{-\frac{x^2}{4B^2t}} dx
\end{equation}
with $A$ and $B$ defined in Appendix \ref{sec:appen_mellin}. Then we find the FRT for normal diffusion to be 
\begin{equation}\label{eq:fpt_normal}
    \mathcal{F}(t\vert x_0)=\frac{B e^{-\frac{A^2}{4t}}}{\sqrt{\pi t}}-B^2 e^{B(A+Bt)} \mbox{erfc}\left( \frac{A+2Bt}{2\sqrt{t}}\right),
\end{equation}
where $\mbox{erfc}(z)=1-2\pi^{-1/2} \int_0^z e^{-t^2} dt$ is the complementary error function \cite{abramowitz1964handbook}. Eq. (\ref{eq:fpt_normal}) can easily be found by performing a Laplace transform of Eq. (\ref{eq:laplace_fpt}) for $\alpha=1$. One may take the limit $B\to\infty$ to find this quantity for normal diffusion with a perfectly reactive boundary, after taking the limit we find the well known result \cite{redner2001guide},
\begin{equation}
    \mathcal{F}(t\vert x_0)=\frac{A}{\sqrt{4 \pi t^3}} e^{-\frac{A^2}{4t}},
\end{equation}
this is also obtained by taking $\alpha \to 1$ in Eq. (\ref{eq:fpt_fox}). 

Performing a similar procedure for the survival probability, Eq. (\ref{eq:survival_exact}) we find the known solution \cite{redner2001guide},
\begin{equation}\label{eq:survival_diffusion}
    Q(t\vert x_0)=e^{B (A+B t)} \mbox{erfc}\left(\frac{A+2 B t}{2 \sqrt{t}}\right)+\mbox{erf}\left(\frac{A}{2 \sqrt{t}}\right).
\end{equation}
Taking the negative of the time derivative of Eq. (\ref{eq:survival_diffusion}) we recover Eq.(\ref{eq:fpt_normal}) as required. One can see that for $B\to \infty$ we obtain the correct solution for a perfectly reactive boundary. 

For the case of $\alpha=1$ for the subordinated local time density, we need to use the following relation \cite{gorenflo1999analytical},
\begin{equation}\label{eq:wright_gaussian}
    \phi\left(-\frac{1}{2},\frac{1}{2};z\right)=\frac{1}{\sqrt{\pi}}e^{-z^2/4},
\end{equation}
which from Eq. (\ref{eq:local_time_exact}) we recover the following result \cite{takacs1995local},
\begin{equation}
    \rho(l,t\vert0)=\sqrt{\frac{K}{\pi t}}e^{-\frac{K}{4t} l^2}.
\end{equation}

\end{appendices}

\bibliography{sn-article}


\end{document}